\documentclass[a4paper,11pt]{article}

\usepackage{float}
\usepackage{ifthen}
\usepackage{natbib}
\usepackage{amsmath}
\usepackage{amssymb}
\usepackage{caption}
\usepackage{tocloft}
\usepackage{booktabs}
\usepackage{geometry}
\usepackage{graphicx}
\usepackage{tabularx}
\usepackage{verbatim}
\usepackage[utf8x]{inputenc}

\graphicspath{{.}}

\newcommand{\bE}{\mathbb{E}}
\newcommand{\bP}{\mathbb{P}}
\newcommand{\dt}{\,\mathrm{d}t}
\newcommand{\QED}{\hfill$\square$}

\newtheorem{lemma}{Lemma}
\newtheorem{remark}{Remark}
\newtheorem{assumption}{Assumption}
\newtheorem{definition}{Definition}
\newtheorem{proposition}{Proposition}

\newcommand{\mystrut}{\vphantom{()}}

\newcommand{\mycaption}[3][]%
   {\ifthenelse{\equal{#1}{}}{\caption{#2}}{\caption[#1]{#2}}%
    \ifthenelse{\equal{#3}{}}{}{{\small #3\mystrut\par}}}

\newcommand{\myfigure}[2][]%
   {\ifthenelse{\equal{#1}{}}%
   {\centerline{\includegraphics[width=\textwidth]{#2}}}%
   {{\centering\small\bf #1\mystrut\par}\vspace{-1ex}%
    \centerline{\includegraphics[width=\textwidth]{#2}}}}

\newcommand{\myminitab}[3][1]%
   {\setlength{\tabcolsep}{0pt}\renewcommand{\arraystretch}{#1}%
   \begin{tabular}{#2}#3\end{tabular}}

\title{\Large\bf The Mathematics of the Relationship between \\ the Default Risk and Yield-to-Maturity of Coupon Bonds}
\author{Sara Cecchetti\thanks{Bank of Italy, Economic Outlook and Monetary Policy Department, Via Nazionale 91, 00184 Roma, Italy. We thank Giuseppe Grande and Giovanna Nappo for their comments and suggestions. The views expressed in the article are those of the authors and do not involve the responsibility of the Bank of Italy.} \thanks{Email: \mbox{sara.cecchetti@bancaditalia.it}.}\and Antonio Di Cesare{\footnotemark[1]} \thanks{Email: \mbox{antonio.dicesare@bancaditalia.it}.}}

\begin{document}

\maketitle

\begin{abstract}
\noindent
The paper analyzes the mathematics of the relationship between the default risk and yield-to-maturity of a coupon bond. It is shown that the yield-to-maturity is driven not only by the default probability and recovery rate of the bond but also by other contractual characteristics of the bond that are not commonly associated with default risk, such as the maturity and coupon rate of the bond. In particular, for given default probability and recovery rate, both the level and slope of the yield-to-maturity term structure depend on the coupon rate, as the higher the coupon rate the higher the yield-to-maturity term structure. In addition, the yield-to-maturity term structure is upward or downward sloping depending on whether the coupon rate is high or low enough. Similar qualitative results also holds for CDS spreads. Consequently, the yield-to-maturity is an indicator that must be used cautiously as a proxy for default risk.  
\end{abstract}

\vfill



\vspace{2em}

\thispagestyle{empty}

\newpage

\cftsetindents{sec}{0pt}{1.5em}
\setlength{\cftbeforesecskip}{1ex}
\renewcommand{\cftsecfont}{}
\renewcommand{\cftsecpagefont}{}
\renewcommand{\cftsecleader}{\cftdotfill{\cftdotsep}}
\renewcommand{\contentsname}{\centerline{\Large{Contents}}}
\tableofcontents

\vspace{0.03\textheight}

\cftsetindents{tab}{0pt}{1.5em}
\setlength{\cftbeforetabskip}{1ex}
\renewcommand{\cfttabfont}{}
\renewcommand{\cfttabpagefont}{}
\renewcommand{\cfttableader}{\cftdotfill{\cftdotsep}}
\renewcommand{\listtablename}{\centerline{\Large{List of Tables}}}
\listoftables

\vspace{0.03\textheight}

\cftsetindents{fig}{0pt}{1.5em}
\setlength{\cftbeforefigskip}{1ex}
\renewcommand{\cftfigfont}{}
\renewcommand{\cftfigpagefont}{}
\renewcommand{\cftfigleader}{\cftdotfill{\cftdotsep}}
\renewcommand{\listfigurename}{\centerline{\Large{List of Figures}}}
\listoffigures

\thispagestyle{empty}
\setcounter{page}{0}

\newpage

\setlength{\baselineskip}{1.3\baselineskip}

\section{Introduction}\label{sec:ytm_intro}

Traditionally, the literature on the valuation of fixed income securities has focussed on the interest rate risk, that is the sensitivity of the bond price to interest rate changes (see, for example, \citealp{Fabozzi:2003,Fabozzi:2007}). The relationship between the bond price and interest rate changes is described using the concept of duration, which also links the bond price sensitivity to interest rate changes to bond characteristics such as the maturity and coupon rate. In the classical framework, longer maturities and lower coupon rates are associated with greater durations, that is with greater price sensitivities to interest rate changes. Being more risky, bonds with longer maturities or lower coupons are generally rewarded with higher yields. This paper shows that these classical relationships can break down completely when the bond is not risk-free,%
\footnote{Throughout the paper, the term ``risk-free'' is used as synonymous of ``default-free''.}
that is when the default probability is not null. In presence of default risk, long-term bonds can have lower yields than short-term bonds and higher coupons can be associated with higher yields. In fact, yields-to-maturity and bond spreads --- the difference between the yield-to-maturity of a defaultable bond and the yield-to-maturity of a risk-free bond --- depend also on other contractual features of the bonds that are not generally regarded as having an impact on default risk, such as the maturity and coupon rate.%
\footnote{In this paper it is assumed that the risk-free rate (i.e., the yield-to-maturity of a risk-free bond) is constant over the considered time horizon, so that the yield-to-maturity of a defaultable bond provides the same information as the bond spread. However, the results presented in this paper are still valid, on a qualitative level, if the risk-free term structure is not flat. This paper leaves aside potential determinants of bond spreads that are different from default risk, such as liquidity, taxation, embedded options, etc. (see, for instance, \citealp{Elton-Gruber-etal:2001}, and \citealp{Avramov-Jostova-etal:2007}).} 
For this reason, yields-to-maturity and bond spreads can be completely misleading when used to infer default probabilities and recovery rates of the bonds and have to be cautiously interpreted as indicators of default risk. The results presented in this paper should be particularly insightful for academics, investors, regulators, supervisors, and policy-makers who use yields-to-maturity and bond spreads as indicators of default risk for bonds.%
\footnote{Just to mention a few international institutions that have recently used bond spreads for their analysis, see \citet{IMF-WEO:2010,IMF-GFSR:2011}, \citet{BIS:2010}, and \citet{ECB-FSR2:2010}.}

More in detail, this paper shows analytically and numerically that the yield-to-maturity of a defaultable bond can vary considerably as a function of the maturity and coupon rate of the bond, for a given level of default probability and recovery rate. In particular, defaultable bonds with greater coupons tends to have higher yields-to-maturity. The intuition behind this result is that greater coupons are associated with less than proportional price increases when the default probability is not null because there is a chance that the coupons are not actually paid due to the default of the issuer. In this case, bond prices are relatively low with respect to the nominal cash flows and determine higher yields-to-maturity. This result implies, for instance, that bonds with higher default probabilities can have lower yields-to-maturity when they have coupon rates that are low enough, and vice versa.

This paper also shows that, in addition to the level, the slope of the yield-to-maturity term structure has to be cautiously interpreted as well, as in general it does not convey enough information to establish if the default probabilities of the bonds are higher in one period than in another. For instance, a downward sloping yield-to-maturity term structure does not necessarily imply that the default probabilities of the bonds are higher on shorter maturities than on longer maturities. This result arises from the fact that the coupon rate of a defaultable bond has an impact also on the slope of the yield-to-maturity term structure. Keeping unchanged the default probability and recovery rate, defaultable bonds with low coupons tend to determine downward sloping yield-to-maturity term structures, while defaultable bonds with high coupons tend to imply increasing yield-to-maturity term structures. The intuition behind these results is that, in case of default, higher coupons determine losses relatively higher for bonds with longer maturities so that those bonds tend to have prices which are relatively lower, and lower prices tend to be associated with higher yields-to-maturity. On the other side, nominal losses in case of default tend to be almost identical for both short-term and long-term bonds when coupons are low. For example, the losses in case of default for zero-coupon bonds are always given only by the amounts lost on the final payoffs of the bonds, as there are no coupons that can be lost. It follows that the prices of defaultable bonds with longer maturities are relatively higher, because the potential losses can happen only in far away time horizons, and the corresponding yields-to-maturity are lower.

The paper is organized as follows. Sections~\ref{sec:ytm_mmI} and~\ref{sec:ytm_mmII} describe the analytical frameworks and contain the main results in discrete and continuos time, respectively. Section~\ref{sec:ytm_numex} provides both some numerical examples and some recent empirical evidence from Greek and Italian government bonds. Section~\ref{sec:ytm_concl} concludes. The formal proofs are collected in the {\appendixname}. The {\appendixname} also shows that most of the results obtained for yields-to-maturity of defaultable coupon bonds also hold for CDS spreads.

\section{The mathematical model I\@: Discrete time}\label{sec:ytm_mmI}

This section introduces a discrete-time model in which there are defaultable coupon bonds maturing at time $T\in\{1,2,\dots\}$.%
\footnote{A list of the symbols used in the paper with their definitions is contained in Table~\ref{tab:ytm_def} in the {\appendixname}.}
The bonds have the same nominal value (set equal to 100) and pay the same periodical coupons $C\geq 0$ (i.e., the coupon rate is $c=C/100$).  For each bond, the probability of being in default at time $t\in\{1,2,\dots\}$ conditional on not having incurred in default in the previous $t-1$ periods is constant and equal to $\lambda\in[0,1]$. The latter assumption can also be expressed in terms of unconditional default probability.
\begin{assumption}\label{ass:ytm_mmI_defpr}
The probability that the default time $\tau$ of a bond with maturity $T\in\{1,2,\dots\}$ is equal to $t\in\{1,\ldots,T\}$ is given by
\begin{equation}
\bP(\tau=t)=
\begin{cases}
\lambda{(1-\lambda)}^{t-1} & \text{if $t\in\{1,\ldots,T\}$} \cr
0 & \text{otherwise},
\end{cases}
\end{equation}
for some constant $\lambda\in[0,1]$, with the convention that
\begin{equation}
\bP(\tau=t)=
\begin{cases}
1 & \text{if $t=1$} \cr
0 & \text{otherwise},
\end{cases}
\end{equation}
when $\lambda=1$.
\end{assumption}
The probability $\lambda$ is a risk-neutral probability,%
\footnote{Alternatively, one can assume that $\lambda$ is the real-world conditional default probability and that investors are risk-neutral.}
implying that the price at time 0 of a bond with maturity $T\in\{1,2,\dots\}$ is equal to the expected value of future payoffs discounted at the risk-free rate,
\begin{equation}\label{eq:ytm_mmI_bond}
\begin{split}
P_{T} &= \sum_{t=1}^{T}C\frac{{(1-\lambda)}^{t}}{{(1+r)}^{t}}+\sum_{t=1}^{T}R\frac{\lambda{(1-\lambda)}^{t-1}}{{(1+r)}^{t}}+100\frac{{(1-\lambda)}^{T}}{{(1+r)}^{T}} \cr
 &= {\left(C+\frac{R\lambda}{1-\lambda}\right)}{\left(\frac{1-\lambda}{r+\lambda}\right)}{\left(1-\frac{{(1-\lambda)}^{T}}{{(1+r)}^{T}}\right)}+100\frac{{(1-\lambda)}^{T}}{{(1+r)}^{T}}
\end{split}
\end{equation}
where $r\geq 0$ is a constant risk-free rate and $R\in[0,100]$ is what the owner of the bond receives in case of default (i.e., the recovery rate is $R/100$). The first term on the right-hand side of the first line of Eq.~\eqref{eq:ytm_mmI_bond} is the present value of the coupons, which are paid only when there is no default before the time they are due; the second term represents the present value of the recovery amount, which is paid when the default happens; the last term is the present value of the nominal value of the bond, which is paid at maturity if there has not been a default previously.

Next, the yield-to-maturity of a bond is formally defined (see, for instance, \citealp{Hull:2008}).
\begin{definition}\label{def:ytm_mmI_ytm}
The yield-to-maturity of a bond with price $P_{T}$ and maturity $T\in\{1,2,\dots\}$, which pays periodical coupons $C\geq 0$ and has nominal value equal to 100, is the unique value $y_{T}$ such that
\begin{equation}\label{eq:ytm_mmI_ytm}
P_{T}=\sum_{t=1}^{T}\frac{C}{{(1+y_{T})}^{t}}+\frac{100}{{(1+y_{T})}^{T}}.
\end{equation}
\end{definition}

In case of a default-free bond (as a Treasury bond) assuming a constant risk-free rate also implies that the yield-to-maturity is exactly equal to the risk-free rate. In fact, it can be immediately noted that Eqs.~\eqref{eq:ytm_mmI_bond} and~\eqref{eq:ytm_mmI_ytm} are equal when $\lambda=0$ and $y_{T}=r$. In this case, the difference between the yield-to-maturity of a defaultable bond and the yield-to-maturity of a default-free bond, the definition of bond spread (or nominal spread), is equivalent to both the zero-volatility spread (\mbox{Z-spread}) and option adjusted spread (OAS). Bond spreads, Z-spreads, and OASs are widely used indicators that measure not only the compensation required by the holders of defaultable bonds for the additional credit risk they bear with respect to the holders of risk-free bonds, but also for differences in liquidity, taxation, and embedded options. In the rest of the paper these additional risk factors are not taken into account and it is assumed that only the default risk component is relevant and determines the spreads.

Another concept that needs to be introduced is that of par yield, which corresponds to the coupon rate for which the bond is quoted at par.
\begin{definition}\label{def:ytm_mmI_paryield}
The par yield $c_{par}$ of a bond with maturity $T\in\{1,2,\dots\}$ is the coupon rate for which the price of the bond is equal to its nominal value:
\begin{equation}\label{eq:ytm_mmI_paryield}
\sum_{t=1}^{T}C_{par}\frac{{(1-\lambda)}^{t}}{{(1+r)}^{t}}+\sum_{t=1}^{T}R\frac{\lambda{(1-\lambda)}^{t-1}}{{(1+r)}^{t}}+100\frac{{(1-\lambda)}^{T}}{{(1+r)}^{T}}=100,
\end{equation}
where $C_{par}=100\,c_{par}$ is the coupon corresponding to the par yield.
\end{definition}
It is worth noting that this definition of par yield is different from the usual definition of par yield $\tilde{c}_{par}=\tilde{C}_{par}/100$ in which the default probabilities are not taken explicitly into account and the future payoffs of a bond are discounted using the appropriate risky rates $\tilde{r}_{1}, \tilde{r}_{2}, \ldots, \tilde{r}_{T}$:
\begin{equation}
100=\sum_{t=1}^{T}\frac{\tilde{C}_{par}}{{(1+\tilde{r}_{t})}^{t}}+\frac{100}{{(1+\tilde{r}_{T})}^{T}}.
\end{equation}
The definition of par yield introduced in Eq.~\eqref{eq:ytm_mmI_paryield} will be extremely useful to prove some interesting results about the relationships between the yield-to-maturity and default probability of the bonds.

As said in the Introduction, this paper aims at showing that the yield-to-maturity must be carefully interpreted as an indicator of credit risk because it depends also on characteristics of the bond that are not related to default risk, such as the maturity $T$ and value of the coupons $C$. As a first step toward attaining this result, notice that from Eqs.~\eqref{eq:ytm_mmI_bond} and~\eqref{eq:ytm_mmI_paryield} it follows that $C_{par}=(100\,r+(100-R)\lambda)/(1-\lambda)$, which implies that the par yield is the same for all bonds, independently of the maturity. This result is formalized in the following lemma.
\begin{lemma}\label{th:ytm_mmI_lemma}
Given $N$ defaultable coupon bonds with maturities $T_{n}\in\{1,2,\ldots\}$, $n=1,\ldots,N$, same nominal value (set equal to 100), same recovery rate $R/100$, and same risk-neutral conditional default probability $\lambda$, and assuming a constant risk-free rate $r$, then there exists a unique par yield $c_{par}$ which is independent of the maturity of the bonds. Such unique value is given by
\begin{equation}\label{eq:ytm_mmI_cpar}
c_{par}=\frac{r+(1-R/100)\lambda}{1-\lambda}.
\end{equation}
\end{lemma}

Eq.~\eqref{eq:ytm_mmI_cpar} shows that the values of the default probability, the recovery rate and the risk-free rate determine the value $c_{par}$ which, however, is independent of the maturity. It is easy to show that the yield-to-maturity of a bond quoted at par with coupons $100\,x$ is equal to $x$.%
\footnote{See the proof of Proposition~\ref{th:ytm_mmI_prop1} in the {\appendixname}.}
It follows that the yield-to-maturity of a bond with coupons equal to $C_{par}=100\,c_{par}$ is equal to the par yield $c_{par}$. Hence, the yield-to-maturity term structure for bonds with coupons equal to $C_{par}$ is flat at the value $c_{par}$.

It follows from Lemma~\ref{th:ytm_mmI_lemma} that bonds with different maturities but exactly the same characteristics in terms of default risk (default probabilities and recovery rates) have the same yield-to-maturity when coupons are equal to $C_{par}$. However, this is an exception, as the following proposition shows.

\begin{proposition}\label{th:ytm_mmI_prop1}
Given a defaultable coupon bond with maturity $T\in\{1,2,\ldots\}$, with a given nominal value (set equal to 100), recovery rate $R/100$, and risk-neutral conditional default probability $\lambda$, and a constant risk-free rate $r$, then the yield-to-maturity of the bond is an increasing function of the coupon $C$ paid by the bond.
\end{proposition}

According to Proposition~\ref{th:ytm_mmI_prop1}, two bonds with same maturity, default probability and recovery rate have different yields-to-maturity if they pay different coupons. The intuition behind this result is that, for defaultable bonds, higher coupons are not fully reflected in higher prices because of the positive probability that a default occurs and the coupons are not paid at all. This argument implies that the left-hand side of Eq.~\eqref{eq:ytm_mmI_ytm}, as derived from Eq.~\eqref{eq:ytm_mmI_bond}, increases relatively less than the right-hand side of the same equation when $C$ becomes larger, so that the yield-to-maturity has to increase as well to balance the equation.

An interesting result which arises from the proof of Proposition~\ref{th:ytm_mmI_prop1} is that when the default probability $\lambda$ or the recovery rate $R/100$ is equal to zero, then the yield-to-maturity of the bond is always equal to the par yield, independently of the value of the other variables, and from the value of the coupon in particular. The rationale behind this result is that, when $\lambda=0$ or $R=0$, Eq.~\eqref{eq:ytm_mmI_ytm} becomes equivalent to Eq.~\eqref{eq:ytm_mmI_bond} by setting $(1+y_{T})=(1+r)/(1-\lambda)=(1+r-\lambda R/100)/(1-\lambda)=(1+c_{par})$. In particular, the yield-to-maturity and par yield are equal to the risk-free rate when $\lambda=0$, that is when the bond is risk-free.

Proposition~\ref{th:ytm_mmI_prop1} shows that the level of the yield-to-maturity term structure depends on the coupon rate. The higher the coupon rate the higher the yield-to-maturity. The following proposition complements the previous result by showing that also the slope of the yield-to-maturity term structure depends on the coupon rate.

\begin{proposition}\label{th:ytm_mmI_prop2}
Given $N$ defaultable coupon bonds with maturities $T_{n}\in\{1,2,\ldots\}$, $n=1,\ldots,N$, same nominal value (set equal to 100), same coupon rate $c$, same recovery rate $R/100$, and same risk-neutral conditional default probability $\lambda$, and a constant risk-free rate $r$, then the yield-to-maturity term structure is upward (downward) sloping when the coupon rate $c$ is higher (lower) than $c_{par}$, that is $y_{T_{1}}\leq y_{T_{2}}\leq\cdots\leq y_{T_{N}}$ ($y_{T_{1}}\geq y_{T_{2}}\geq\cdots\geq y_{T_{N}}$). The term structure of bond prices has a similar behaviour, with prices decreasing (increasing), as a function of maturity, when $c>c_{par}$ ($c<c_{par}$), that is $P_{T_{1}}\geq P_{T_{2}}\geq\cdots\geq P_{T_{N}}$ ($P_{T_{1}}\leq P_{T_{2}}\leq\cdots\leq P_{T_{N}}$).
\end{proposition}

The intuition behind the results stated in Proposition~\ref{th:ytm_mmI_prop2} is that higher coupons determine losses relatively higher for bonds with longer maturities in case of default, so that these bonds tend to have prices which are relatively lower than the prices of bonds with shorter maturities. Lower prices, in turn, tend to be associated with higher yields-to-maturity. When coupons are low, on the other side, nominal losses in case of default tend to be similar for both short-term and long-term bonds. For example, the losses in case of default for zero-coupon bonds are given only by what is lost on the final payoffs of the bonds. It follows that the prices of bonds with longer maturities are relatively higher, because for those bonds the losses would only happen in far away time horizons. As a consequence, the yields-to-maturity of long-term bonds tend to be lower than the yields-to-maturity of short-term bonds.

Proposition~\ref{th:ytm_mmI_prop2} shows that a variation of the coupon rate determines a variation of the yield-to-maturity in the same direction. However, the magnitude of the yield-to-maturity change is smaller than that of the coupon rate.

\begin{remark}\label{th:ytm_mmI_remark}
\rm
For a defaultable coupon bond with maturity $T\in\{1,2,\ldots\}$, coupon $C$, and yield-to-maturity $y_{T}$, the following relationship holds between the coupon variation $\Delta=C-C_{par}$ and related yield-to-maturity variation $\delta_T=y_{T}-c_{par}$, with $c_{par}=C_{par}/100$:
\begin{equation}\label{eq:ytm_mmI_remark}
|\delta_T|<|\Delta|/100.
\end{equation}
\end{remark}

\section{The mathematical model II\@: Continuous time}\label{sec:ytm_mmII}

The previous results can be generalized to a continuous-time model that allows the bonds to have any maturity $T>0$ and defaults to happen at any time $\tau>0$. Hence, both maturities and default times are not restricted to be integer numbers. 

The continuos-time model which is used is the simplest version of a default intensity model in which the default is defined as the first arrival time of a Poisson process with a constant mean arrival rate (intensity) $\lambda$. Roughly speaking, the default probability over a small time period of length $\Theta$, conditional to the survival up to the beginning of that period, is approximated by $\lambda\Theta$ (see, for example, \citealp{Giesecke:2004}, and references therein). This idea is formalized in the following statement.
\begin{assumption}\label{ass:ytm_mmII_defpr}
The time-to-default $\tau$ of a bond is exponentially distributed, that is the probability of surviving for at least $t$ years is given by 
\begin{equation}
\bP(\tau>t)=e^{-\lambda t},
\end{equation}
for some $\lambda\geq 0$.%
\footnote{Natural extensions of this model include the cases in which the intensity is modeled as a piecewise constant, a deterministic function, or a random process. See, for example, \citet{Duffie-Singleton:1999,Duffie-Singleton:2003} and \citet{Lando:2004}.}
\end{assumption}
It can be shown that in this framework the expected time-to-default is $1/\lambda$. Loosely speaking, the default arrival rate $\lambda$ can thus also be interpreted as the expected number of defaults per unit of time.

As in Section~\ref{sec:ytm_mmI}, both the risk-free interest rate $r\geq 0$ and coupon $C\geq 0$ are assumed to be constant. However, in the current framework both $r$ and $C$ are also assumed to be paid continuously over time, that is the amounts $rX$ and $C$ are paid per unit of time, where $X$ is the nominal value of a default-free bond. The payoffs of a defaultable coupon bond with maturity $T$ and nominal value equal to $100$, under the further assumption that the amount which is recovered in case of default if $R\in[0,100]$, are thus given by:
\begin{itemize}
\item[\emph{i})] the coupon $C$, corresponding to a claim which is paid continuously until maturity or until a default occurs;
\item[\emph{ii})] the recovery amount $R$, which is paid at the default time $\tau$ if a default occurs;
\item[\emph{iii})] the nominal value of the bond $100$, which is paid at maturity if no default occurs.
\end{itemize}
As shown by \citet[ch. 5]{Lando:2004}, assuming that the default arrival rate $\lambda$ defines risk-neutral default probabilities, the price of a defaultable coupon bond at time 0 is given by the expected payoffs discounted at the risk-free rate and it is equal to
\begin{equation}\label{eq:ytm_mmII_bond}
\begin{split}
P_{T} &= \bE{\left[\int_{0}^{T}C\,e^{-rt}\mathbf{1}_{\{\tau>t\}}\dt+R\,e^{-r\tau}\mathbf{1}_{\{\tau\leq T\}}+100\,e^{-rT}\mathbf{1}_{\{\tau>T\}}\right]} \cr
 &= C\!\int_{0}^{T}e^{-(r+\lambda) t} \dt +R\lambda\!\int_{0}^{T}e^{-(r+\lambda)t} \dt +100\,e^{-(r+\lambda)T} \cr
 &= \frac{C+R\lambda}{r+\lambda}{\left(1-e^{-(r+\lambda)T}\right)}+100\,e^{-(r+\lambda)T},
\end{split}
\end{equation}
where
\begin{equation}
\mathbf{1}_{\{A\}}=
\begin{cases}
1 & \text{if the event $A$ occurs} \cr
0 & \text{otherwise}.
\end{cases}
\end{equation}

Similarly to the discrete-time case, the yield-to-maturity of a bond can be defined in a continuous-time framework as well.
\begin{definition}\label{def:ytm_mmII_ytm}
The yield-to-maturity of a bond which continuously pays a coupon $C$, has nominal value 100, and maturity $T>0$, is the unique value $y_{T}$ such that  
\begin{equation}\label{eq:ytm_mmII_ytm}
P_{T}=C\!\int_{0}^{T}e^{-y_{T}t}\dt+100\,e^{-y_{T}T}.
\end{equation}
\end{definition}

The definition of par yield becomes
\begin{definition}\label{def:ytm_mmII_paryield}
The par yield $c_{par}$ of a bond with maturity $T>0$ is the coupon rate for which the price of the bond is equal to its nominal value,
\begin{equation}\label{eq:ytm_mmII_paryield}
C_{par}\!\int_{0}^{T}\,e^{-(r+\lambda) t}\dt+R\lambda\!\int_{0}^{T}e^{-(r+\lambda)t}\dt+100\,e^{-(r+\lambda)T}=100,
\end{equation}
where $C_{par}=100\,c_{par}$ is the coupon corresponding to the par yield.
\end{definition}

Using Eqs.~\eqref{eq:ytm_mmII_bond} and~\eqref{eq:ytm_mmII_paryield} it can be verified that $C_{par}=100\,r+(100-R)\lambda$. This result immediately shows that the par yield is independent of the maturity of the bond and a statement analogous to that reported in Lemma~\ref{th:ytm_mmI_lemma} follows.

\begin{lemma}\label{th:ytm_mmII_lemma}
Given $N$ defaultable coupon bonds with maturities $T_{n}>0$, $n=1,\ldots,N$, same nominal value (set equal to 100), same recovery rate $R/100$, and same risk-neutral default arrival rate $\lambda$, and assuming a constant risk-free rate $r$, then there exists a unique par yield $c_{par}$ which is indipendent of the maturity of the bonds. Such unique value is given by
\begin{equation}\label{eq:ytm_mmII_cpar}
c_{par}=r+(1-R/100)\lambda.
\end{equation}
\end{lemma}

The {\appendixname} shows that also the results stated in Proposition~\ref{th:ytm_mmI_prop1}, Proposition~\ref{th:ytm_mmI_prop2}, and Remark~\ref{th:ytm_mmI_remark} can be generalized in the continuous-time model.

\begin{proposition}\label{th:ytm_mmII_prop1}
Given a defaultable coupon bond with maturity $T>0$, with a given nominal value (set equal to 100), recovery rate $R/100$, and risk-neutral default arrival rate $\lambda$, and a constant risk-free rate $r$, then the yield-to-maturity of the bond is an increasing function of the coupon $C$ paid by the bond.
\end{proposition}

\begin{proposition}\label{th:ytm_mmII_prop2}
Given $N$ defaultable coupon bonds with maturities $T_{n}>0$, $n=1,\ldots,N$, same nominal value (set equal to 100), same coupon rate $c$, same recovery rate $R/100$, and same risk-neutral default arrival rate $\lambda$, and a constant risk-free rate $r$, then the yield-to-maturity term structure is upward (downward) sloping when the coupon rate $c$ is higher (lower) than $c_{par}$, that is $y_{T_{1}}\leq y_{T_{2}}\leq\cdots\leq y_{T_{N}}$ ($y_{T_{1}}\geq y_{T_{2}}\geq\cdots\geq y_{T_{N}}$). The term structure of bond prices has a similar behaviour, with prices decreasing (increasing), as a function of maturity, when $c>c_{par}$ ($c<c_{par}$), that is $P_{T_{1}}\geq P_{T_{2}}\geq\cdots\geq P_{T_{N}}$ ($P_{T_{1}}\leq P_{T_{2}}\leq\cdots\leq P_{T_{N}}$).
\end{proposition}

\begin{remark}\label{th:ytm_mmII_remark}
\rm
For a coupon bond with maturity $T>0$, the following relationship holds between the coupon variation $\Delta$ with respect to $C_{par}$ and related yield-to-maturity increment $\delta_T$ with respect to $c_{par}=C_{par}/100$:
\begin{equation}\label{eq:ytm_mmII_remark}
|\delta_T|<|\Delta|/100.
\end{equation}
\end{remark}

Summarizing, all the results presented for the discrete-time case can be extended to the continuous-time framework. In particular, both the level and slope of the yield-to-maturity curve depend on the value of coupons paid by the bonds.

\section{Numerical examples}\label{sec:ytm_numex}

This section provides a few numerical examples that highlight the empirical relevance of the previous theoretical results. The calculations are done using the equations described in the discrete-time case because the conventions used therein are closer to the market practice, but it is not surprising that any example can be easily replicated in the continuos-case framework given the analogies between the two cases.

\begin{figure}[H]
\myfigure[Panel A\@: Conditional default probability $\lambda=1\%$]{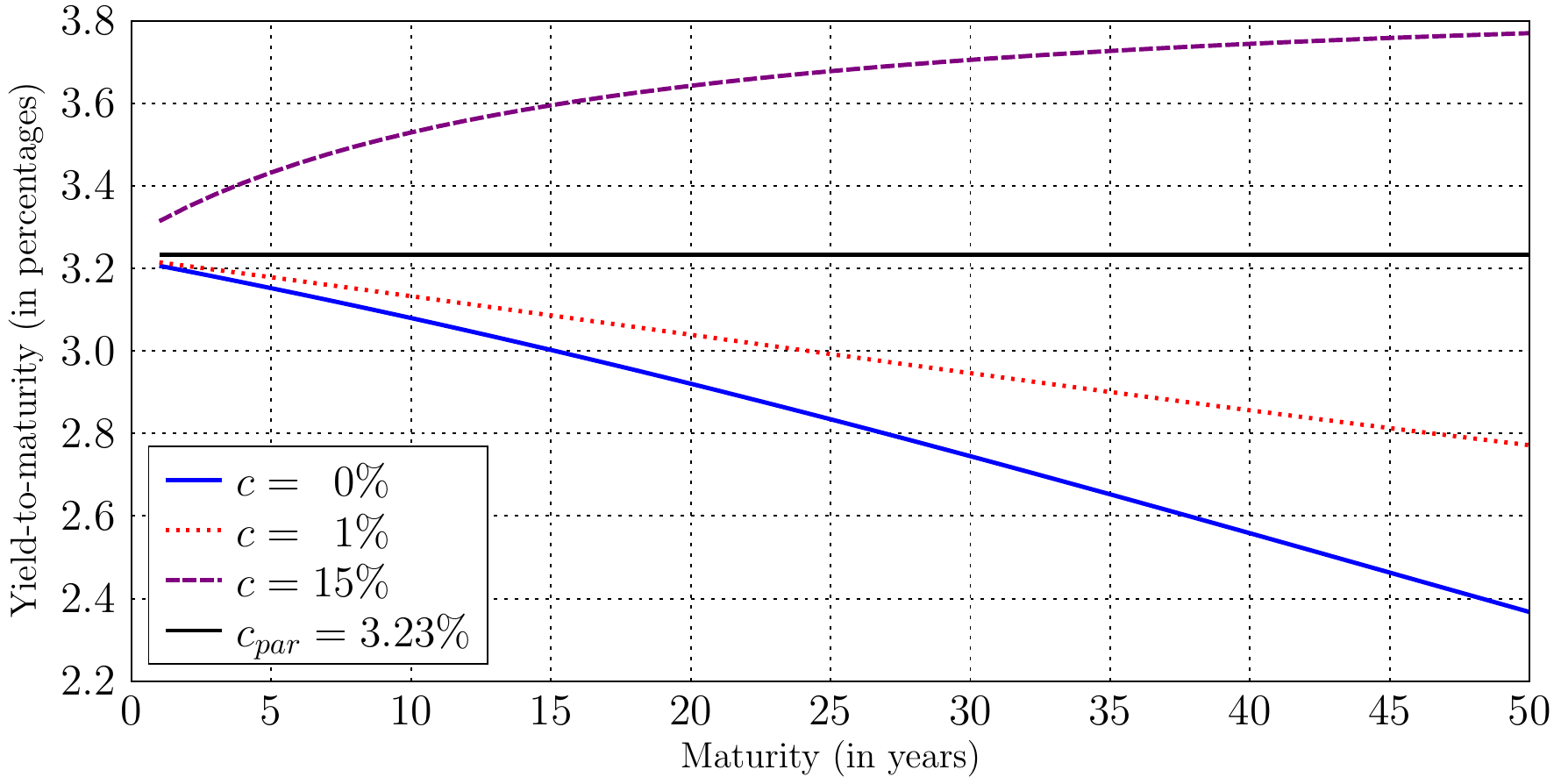}
\myfigure[Panel B\@: Conditional default probability $\lambda=10\%$]{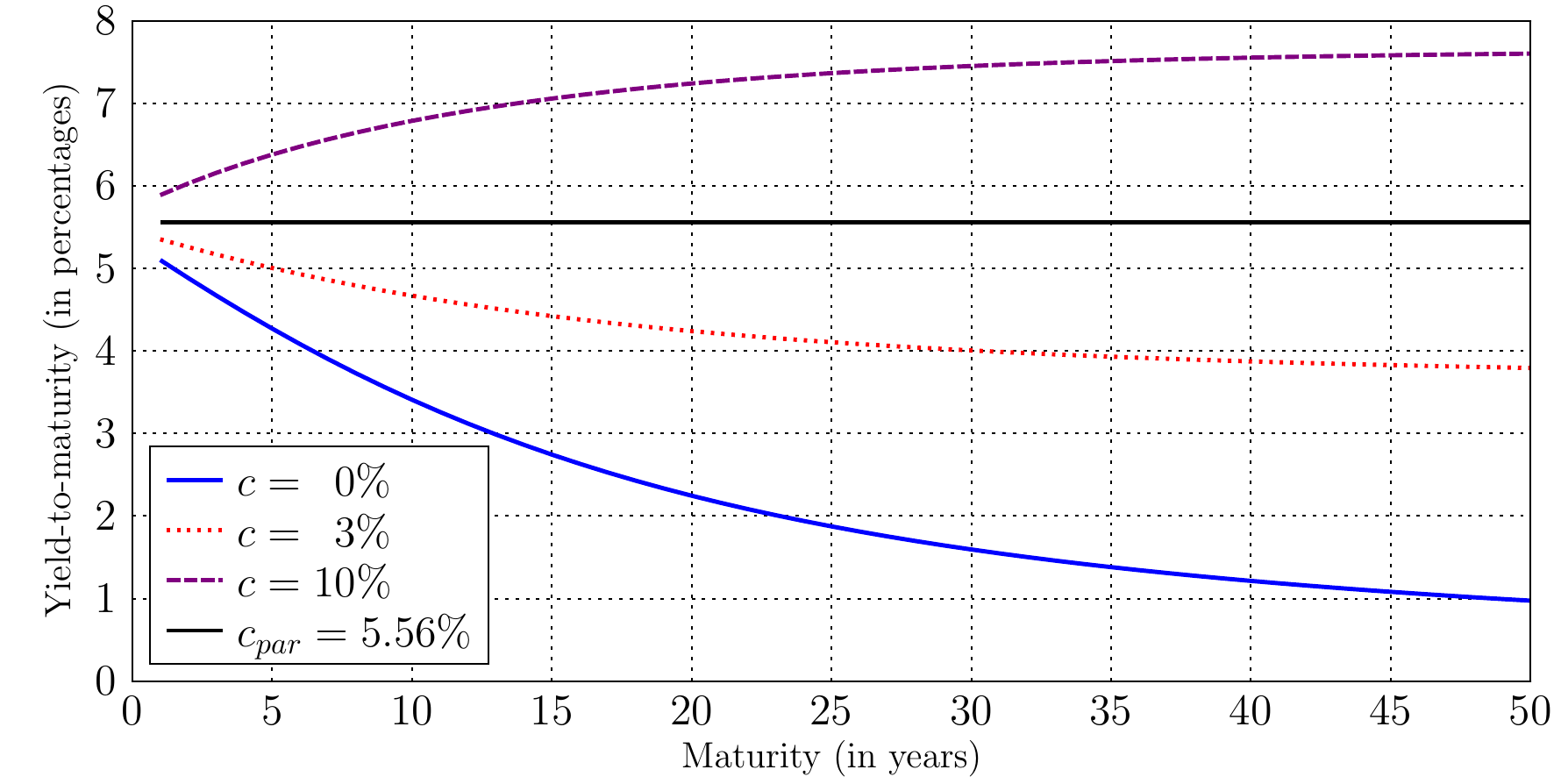}
\mycaption{Yield-to-maturity term structure for different levels of a constant conditional default probability $\lambda$ and coupon rate $c$.}{The remaining assumptions are: risk-fre rate $r=3\%$ and recovery amount $R=80$. The flat black lines represents the par yield $c_{par}$ calculated using Eq.~\eqref{eq:ytm_mmI_cpar} on page~\pageref{eq:ytm_mmI_cpar}.}\label{fig:ytm_fig1}
\end{figure}

In a first set of examples, it is assumed that $r=3\%$ and $R=80$. Figure~\ref{fig:ytm_fig1} shows the yield-to-maturity term structures for two values of $\lambda$ (1\% in Panel A and 10\% in Panel B) under several assumptions on the coupon rates of the bonds. The figure highlights four interesting features of the yield-to-maturity term structure that reflect the theoretical results presented above. First, the yield-to-maturity is an increasing function of the coupon rate, for given maturity. For instance, when $\lambda=10\%$, a 10-year bond can have a yield-to-maturity of either $3.41\%$ or $6.79\%$ depending on whether it is a zero-coupon bond or a bond with a $10\%$ coupon rate. As another striking example, a 10-year bond with a $15\%$ coupon rate has a yield-to-maturity of $3.53\%$ when $\lambda=1\%$, but the yield-to-maturity decreases to $3.41\%$ when the default probability $\lambda$ is ten times greater and the bond is zero-coupon. These results are due to the fact that, when there is a positive default probability, greater coupons are associated with less than proportional price increases, because there is a probability, linked to the default likelihood, that the coupons are not actually paid off. In this case, bond prices are relatively low with respect to the nominal cash flows and yields-to-maturity, which are calculated on nominal cash flows, are relatively higher. It is remarkable that the greater bond risk and associated yield-to-maturity are not determined by a larger default probability or a smaller recovery rate, but only by other characteristics of the bond (maturity and coupon rate). It follows that it is possible to have different yields-to-maturity for given default probabilities and recovery rates, and vice versa, as emphasized by these examples.

Second, for a given coupon rate, the yield-to-maturity is a monotonic function of the bond maturity. Under the previous assumptions on the risk-free rate and recovery rate, using Eq.~\eqref{eq:ytm_mmI_cpar}, one has $c_{par}=3.23\%$ when $\lambda=1\%$ and $c_{par}=5.56\%$ if $\lambda=10\%$. Figure~\ref{fig:ytm_fig1} shows that the yield-to-maturity term structure is downward sloping when the coupon rate is lower than $3.23\%$ or $5.56\%$, according to the value of $\lambda$, and upward sloping otherwise. When coupons are high, a large share of the losses in case of default is represented by the coupons that are not longer paid off. The losses related to the missing payment of the coupons are greater the longer the maturity of the bonds, because more coupons may get lost, so that long-term bonds tend to have lower prices and higher yields-to-maturity than short-term bonds. On the other hand, when coupons are low a larger share of the losses in case of default is represented by the loss on the nominal amount of the bonds, which is the same for all bonds. Because the recovery rate is assumed to be fixed, the present value of the loss on the nominal amount of the bonds is smaller the greater the maturity of the bonds. As a consequence, the prices of bonds with low coupons tend to be higher when their maturity is longer and yields-to-maturity of long-term bonds tend to be lower than yields-to-maturity of short-term bonds.

Third, another problem that arises when yields-to-maturity are used to infer information about the default probability of the bonds is due to the role of the actual residual life of the bonds. Bonds with higher default probabilities have lower actual residual life, and this property can be reflected in prices which are relatively higher than for bonds with lower default probability (and higher actual residual life), because the cash flows are discounted for shorter time periods. Bonds with higher default probabilities, and recovery rates sufficiently high, can thus have lower yields-to-maturity than bonds with lower default probabilities, even if they have the same coupon rates. In the examples provided by Figure~\ref{fig:ytm_fig1}, one can notice that the yields-to-maturity for zero-coupon bonds with $\lambda=1\%$ and maturity equal or greater than 13 years are greater than the yields-to-maturity for zero-coupon bonds with corresponding maturity and $\lambda=10\%$.

Fourth, it is worth noticing that the yields-to-maturity of bonds with low coupons and maturity long enough can be even lower than the risk-free rate, in spite of high default probabilities. Although this result should not be surprising given the arguments mentioned above, it is useful to restate that this happens because there is a positive probability that the bonds are paid well before the maturity, because of default. Given the characteristics of the bonds in terms of coupon rates and recovery rates, the losses on the coupons which are lost and the nominal amount which is not paid back are low, and more than compensated by the fact that the recovery amount is received by the investors much more in advance than what would be the case if the bond were reimbursed at maturity. Finally, it can be observed that if the risk-free rate were lower, the gains in terms of expected value coming from the anticipated repayment could not be sufficient to compensate the losses. For example, when $r=3\%$ and $\lambda=10\%$ the yield-to-maturity of a zero-coupon bond with maturity 15 years is $2.74\%$, which is lower than the risk-free rate, but the yield-to-maturity for the same bond when $r=2\%$ is $2.24\%$, which is greater than the risk-free rate.

The previous examples were based on the theoretical results presented in Section~\ref{sec:ytm_mmI} and extended in Section~\ref{sec:ytm_mmII}. In those sections, because of analytical tractability, several simplifying assumptions were made. However, one may expect that, because of continuity, the gist of the results presented above should still hold true if the underlying assumptions are only slightly modified. In order to show that this is indeed the case, a few numerical examples are now presented in which the assumption that the default probability $\lambda$ is constant is removed. In this case, the yield-to-maturity can still be calculated using Eq.~\eqref{eq:ytm_mmI_ytm} and taking into account that Eq.~\eqref{eq:ytm_mmI_bond} is modified as follows:
\begin{equation}
P_{T}=\sum_{t=1}^{T}C\frac{\prod_{n=1}^{t}(1-\lambda_{n})}{{(1+r)}^{t}}+\sum_{t=1}^{T}R\frac{\lambda_{t}\prod_{n=1}^{t-1}(1-\lambda_{n})}{{(1+r)}^{t}}+100\frac{\prod_{n=1}^{T}(1-\lambda_{n})}{{(1+r)}^{T}},
\end{equation}
where $\lambda_{t}$ is the default probability at time $t$ conditional on not having defaulted before.

\begin{figure}[H]
\myfigure[Panel A\@: Conditional default probability \par $\lambda_{t}\in\{10.0\%,10.5\%,11.0,\ldots,34.5\%\}$]{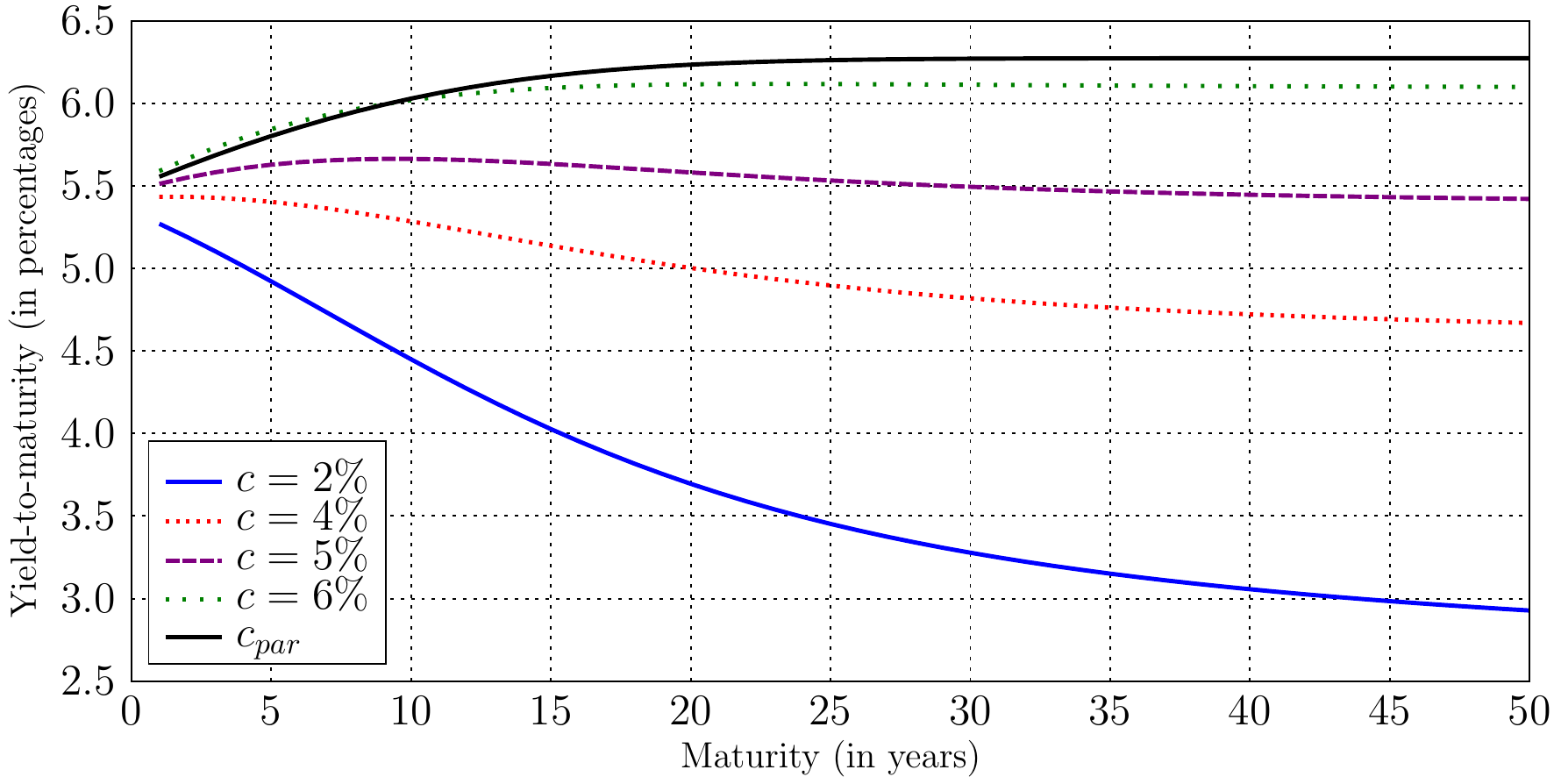}
\myfigure[Panel B\@: Conditional default probability \par $\lambda_{t}\in\{10.0\%,9.8\%,9.6,\ldots,0.2\%\}$]{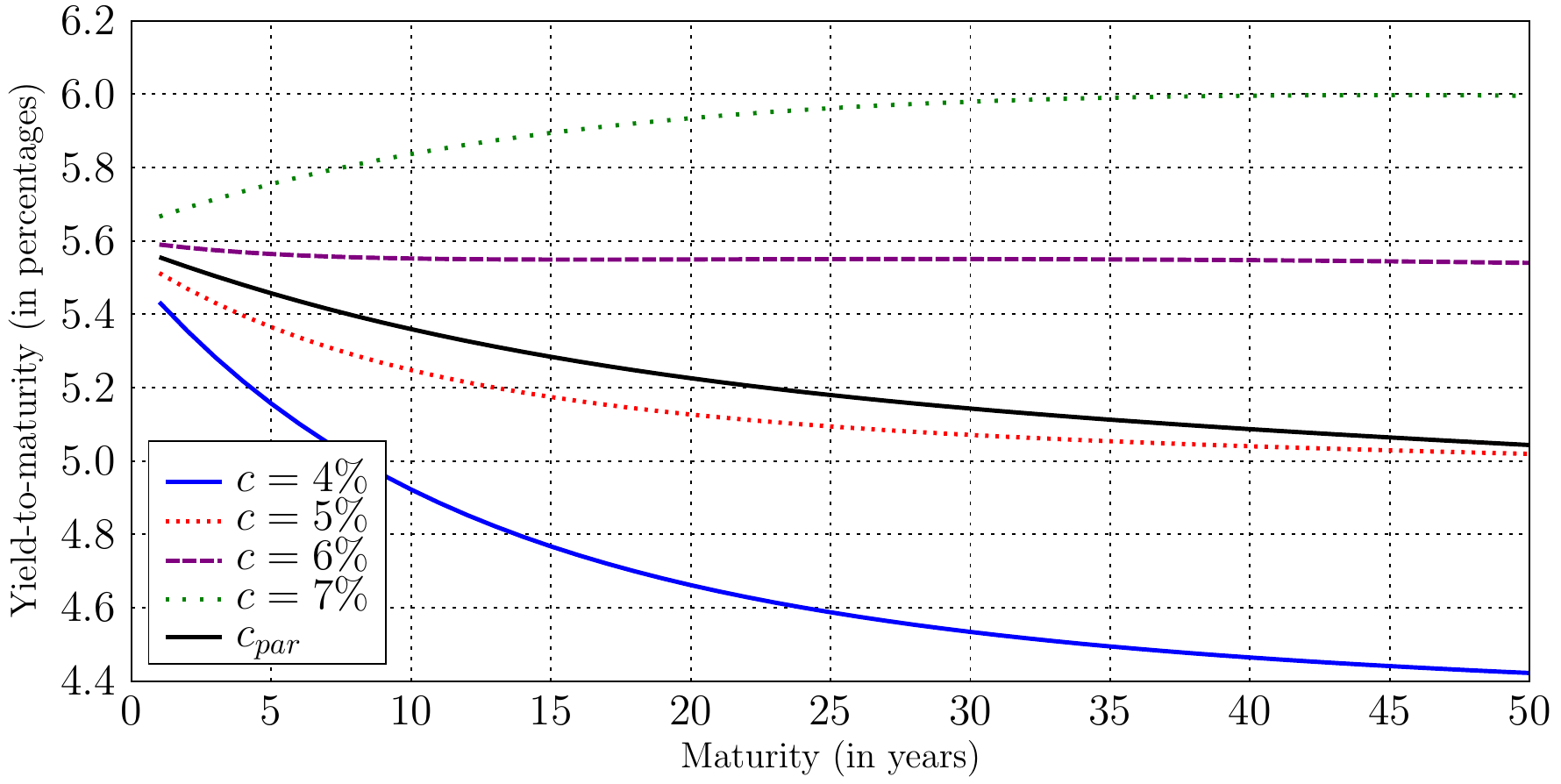}
\mycaption{Yield-to-maturity term structure for different levels of a time-varying conditional default probability $\lambda_{t}$ and coupon rate $c$.}{The remaining assumptions are: risk-free rate $r=3\%$ and recovery amount $R=80$. The flat black lines represents the par yield $c_{par}$ calculated using Eq.~\eqref{eq:ytm_mmI_cpar} on page~\pageref{eq:ytm_mmI_cpar}.}\label{fig:ytm_fig2}
\end{figure}

In this second set of examples it is still assumed that $r=3\%$ and $R=80$. Figure~\ref{fig:ytm_fig2} shows the yield-to-maturity term structures for the cases in which $\lambda_{t}$ is linearly increasing between the years 1 and 50 (from 10.0\% to 34.5\%, Panel A) and linearly decreasing over the same time horizon (from 10.0\% to 0.2\%, Panel B). There are three interesting results that appear from these calculations. First, the par yield is no longer constant. It is increasing when the default probability is increasing as well and decreasing in the other case, although in both conditions par yield changes appears to be much smaller than changes in default probability. For instance, Panel A shows that the par yield is almost constant between 20 and 50 years, although the default probability continues to rise considerably over that time horizon. Second, the level of the yield-to-maturity, for given maturity, can still vary considerably as a function of the coupon rate. As an example, a 10-year bond in Panel B can have a yield-to-maturity from 4.92\% to 5.84\%, depending on the coupon rate that can vary between 4\% and 7\%. Finally, the shape of the yield-to-maturity term structure looks still completely unrelated with the shape of the default probability term structure. The yield-to-maturity term structure can be increasing or decreasing irrespective of whether the default probabilities are increasing or decreasing. Once again, high and low coupon rates tend to be associated with increasing and decreasing yield-to-maturity term structures, respectively. Even a humped yield-to-maturity term structure is possible, although the default probability term structure is linear.

Overall, the numerical examples presented in this section reinforce the previous theoretical findings and confirm that the yield-to-maturity of a bond can be very weakly linked to the default probability of the issuer.

\begin{figure}[H]
\myfigure[Panel A\@: Yields-to-maturity and conditional default probabilities \par of Italian benchmark government bonds]{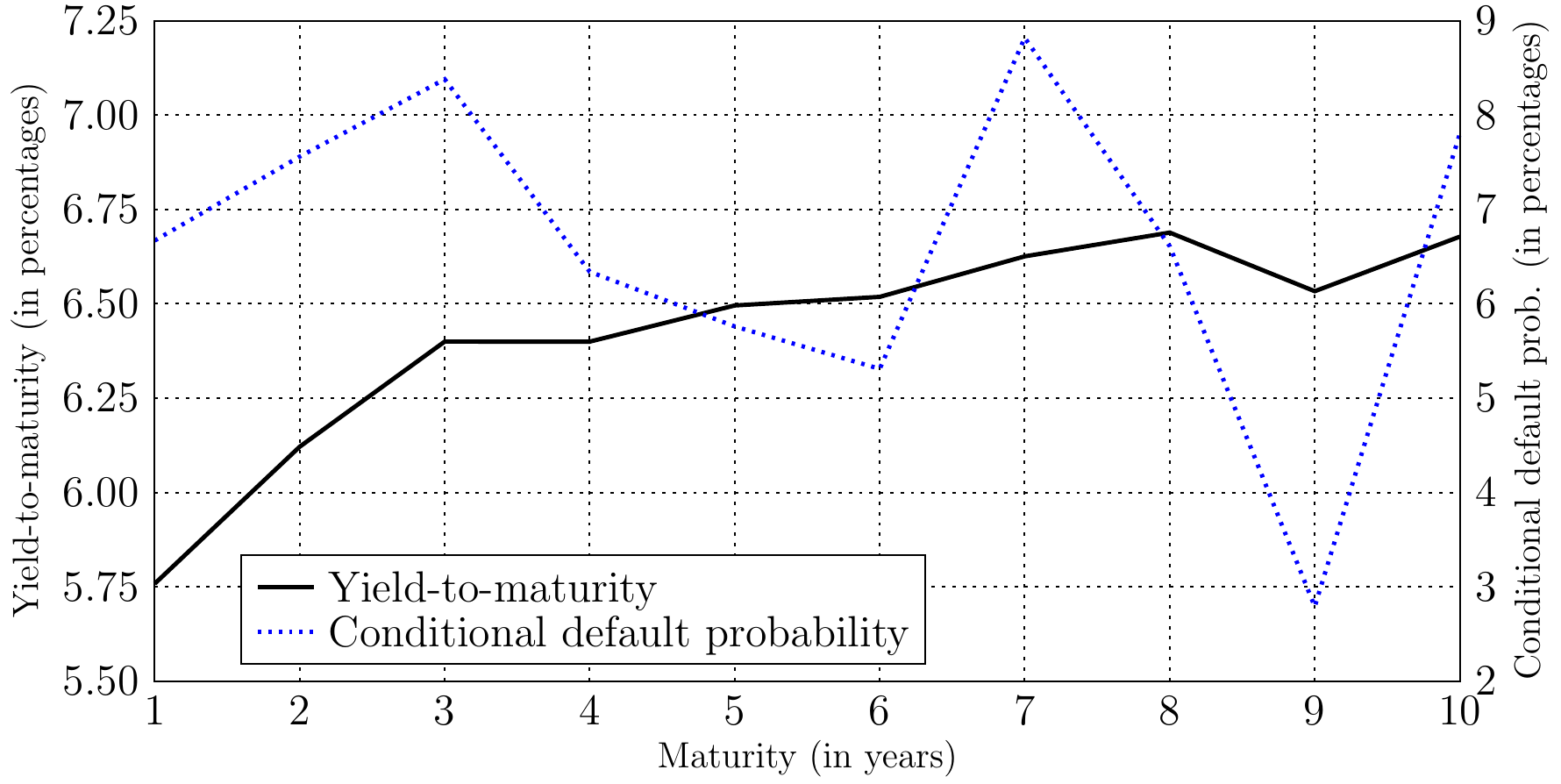}
\myfigure[Panel B\@: Yields-to-maturity and conditional default probabilities \par of Greek benchmark government bonds]{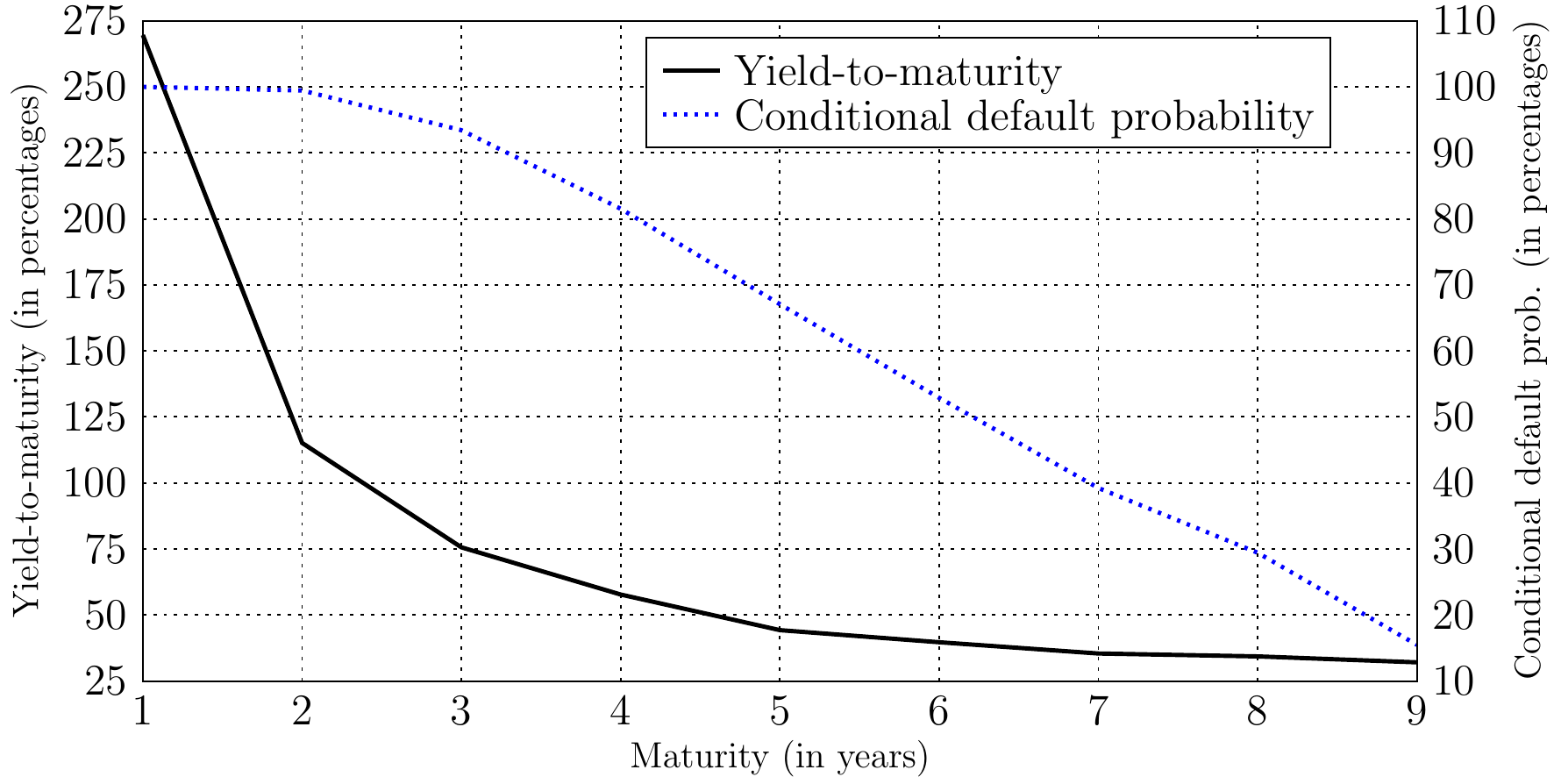}
\mycaption{Yield-to-maturity term structures and conditional default probabilities of Italian and Greek benchmark government bonds}{The conditional default probabilities are calculated using a generalization of Eq.~\eqref{eq:ytm_mmII_bond} in which the default intensity is assumed to be piecewise constant for each year. The recovery amount is set to 40. Italian and Greek benchmark bond prices are taken from Bloomberg. We use the curve on German government bonds, as estimated by Datastream, for the risk-free rates. Data refer to November 18, 2011.}\label{fig:ytm_fig3}
\end{figure}

As an empirical exercise, Figure~\ref{fig:ytm_fig3} shows the yields-to-maturity on November 18, 2011, of Italian and Greek benchmark government bonds on the maturities from one to ten years (nine years for Greece), and the corresponding conditional default probabilities for each year. It is worth noting that both yield curves are (approximately) monotone, with the Italian curve being increasing and the Greek curve being decreasing. According to the common interpretation, these shapes should be coherent with default probabilities that are either increasing (for Italy) or decreasing (for Greece). However, using a generalization of Eq.~\eqref{eq:ytm_mmII_bond} in which the default intensity is assumed to be piecewise constant for each year, it can be shown that this is not necessarily the case, as highlighted in the preceding sections. For Italy, for instance, the conditional default probabilities are increasing in the first three years, but are then decreasing for the maturities between four and six years, notwithstanding the fact that the yield curve continues to increase for these maturities. As for Greece, the yield curve is steeply decreasing, but the conditional default probabilities stand at about 100 per cent for the first four years, and start decreasing only afterwards.

\section{Conclusion}\label{sec:ytm_concl}

The theoretical results and numerical examples provided in this paper show that the yield-to-maturity of a bond depends not only on the default probability and recovery rate (and on the risk-free rate, of course), but also on the bond contractual features, such as the maturity and value of the coupons. For given default probability and recovery rate, both the level and slope of the yield-to-maturity term structure depend on the value of the coupons. As a consequence, yields-to-maturity must be used carefully to infer information about default probabilities, even leaving aside other potential determinants of their levels (interest rate risk, liquidity risk, taxation, embedded options, etc.). Whenever one is interested in having a proxy for the default probability of bond issuers implied in bond prices, it would be better to estimate it using a formal model for pricing defaultable bond than using the shortcut of the yield-to-maturity. The harder work will probably be greatly compensated.

\section*{\appendixname}
\addcontentsline{toc}{section}{\appendixname}
\renewcommand{\theequation}{A.\arabic{subsection}.\arabic{equation}}
\renewcommand{\thesubsection}{A.\arabic{subsection}}

\setcounter{subsection}{0}

\subsection{Proof of Proposition~\ref{th:ytm_mmI_prop1}}
\setcounter{equation}{0}

The first step of the proof shows that coupons higher than $C_{par}$ imply yields-to-maturity higher than the par yield, and vice versa. To this end, notice that the following equation holds for any bond with coupons $C=100\,c$
\begin{equation}\label{eq:ytm_mmI_generic}
100=\sum_{t=1}^{T}\frac{C}{{(1+c)}^{t}}+\frac{100}{{(1+c)}^{T}},
\end{equation}
given that
\begin{equation}
\begin{split}
\sum_{t=1}^{T}\frac{C}{{(1+c)}^{t}}+\frac{100}{{(1+c)}^{T}} &= 100\,c \sum_{t=1}^{T}\frac{1}{{(1+c)}^{t}}+\frac{100}{{(1+c)}^{T}} \cr
 &= 100\,c \frac{\frac{1}{1+c}-\frac{1}{{(1+c)}^{T+1}}}{1-\frac{1}{1+c}}+\frac{100}{{(1+c)}^{T}} \cr
 &= 100\,c \frac{1-\frac{1}{{(1+c)}^{T}}}{c}+\frac{100}{{(1+c)}^{T}} \cr
 &= 100.
\end{split}
\end{equation}
In particular, Eq.~\eqref{eq:ytm_mmI_generic} holds when $C=C_{par}=100\,c_{par}$, so that
\begin{equation}\label{eq:ytm_mmI_generic_paryield}
100=\sum_{t=1}^{T}\frac{C_{par}}{{(1+c_{par})}^{t}}+\frac{100}{{(1+c_{par})}^{T}}.
\end{equation}
Lemma 1 and Eq.~\eqref{eq:ytm_mmI_generic_paryield} imply that the yield-to-maturity of any bond with maturity $T$ is equal to the par yield when the coupons are equal to $C_{par}$. Moreover, it follows from Eqs.~\eqref{eq:ytm_mmI_ytm} and~\eqref{eq:ytm_mmI_paryield} that when a bond has a coupon which differs from $C_{par}$ by $\Delta$, the variation $\delta$ of its yield-to-maturity with respect to $c_{par}$ is implicitly defined by
\begin{equation}\label{eq:ytm_mmI_proof_prop1_eq1}
\begin{split}
\sum_{t=1}^{T}(C_{par}+\Delta)\frac{{(1-\lambda)}^{t}}{{(1+r)}^{t}}+\sum_{t=1}^{T} & R\frac{\lambda{(1-\lambda)}^{t-1}}{{(1+r)}^{t}}+100\frac{{(1-\lambda)}^{T}}{{(1+r)}^{T}}= \cr
 & =\sum_{t=1}^{T}\frac{C_{par}+\Delta}{{(1+c_{par}+\delta)}^{t}}+\frac{100}{{(1+c_{par}+\delta)}^{T}},
\end{split}
\end{equation}
or
\begin{equation}\label{eq:ytm_mmI_proof_prop1_eq2}
100+\Delta\sum_{t=1}^{T}\frac{{(1-\lambda)}^{t}}{{(1+r)}^{t}}=\sum_{t=1}^{T}\frac{C_{par}+\Delta}{{(1+c_{par}+\delta)}^{t}}+\frac{100}{{(1+c_{par}+\delta)}^{T}}.
\end{equation}
When $\delta=0$, the right-hand side of Eq.~\eqref{eq:ytm_mmI_proof_prop1_eq2} becomes, using Eqs.~\eqref{eq:ytm_mmI_generic_paryield} and~\eqref{eq:ytm_mmI_cpar},
\begin{equation}\label{eq:ytm_mmI_proof_prop1_eq3}
\begin{split}
\sum_{t=1}^{T}\frac{C_{par}+\Delta}{{(1+c_{par})}^{t}}+\frac{100}{{(1+c_{par})}^{T}} &= 100+\Delta\sum_{t=1}^{T}\frac{1}{{(1+c_{par})}^{t}} \cr
 &= 100+\Delta\sum_{t=1}^{T}{\left(\frac{1-\lambda}{1+r-R\lambda/100}\right)}^{t},
\end{split}
\end{equation}
which is greater (smaller) than the left-hand side of Eq.~\eqref{eq:ytm_mmI_proof_prop1_eq2} when $\Delta$ is positive (negative). It follows that $\delta$ has to be positive (negative) when $\Delta$ is positive (negative) for Eq.~\eqref{eq:ytm_mmI_proof_prop1_eq2} to be satisfied. It is worth noting that Eq.~\eqref{eq:ytm_mmI_proof_prop1_eq3} is always satisfied for $\delta=0$ when $\lambda=0$ or $R=0$; the yield-to-maturity of a bond with zero default probability or zero recovery rate is always equal to the par yield, whatever the coupon rate. \newline
The second step of the proof shows that the variation $\delta$ of the yield-to-maturity is always smaller than 
\begin{equation}\label{eq:ytm_mmI_proof_prop1_eq4}
\delta_{\sup}=\frac{R\lambda/100}{1-\lambda},
\end{equation}
which means that the yield-to-maturity of a bond has an upper bound $c_{par}+\delta_{\sup}$ that does not depend on the value of the coupon. To prove this result, notice that when $\delta=\delta_{\sup}$ the right-hand side of Eq.~\eqref{eq:ytm_mmI_proof_prop1_eq1} is equal to
\begin{equation}\label{eq:ytm_mmI_proof_prop1_eq5}
\begin{split}
\sum_{t=1}^{T} & \frac{C_{par}+\Delta}{{(1+c_{par}+\delta_{\sup})}^{t}}+\frac{100}{{(1+c_{par}+\delta_{\sup})}^{T}}= \cr
 &= \sum_{t=1}^{T}\frac{C_{par}+\Delta}{{\left(1+\frac{r+(1-R/100)\lambda}{1-\lambda}+\frac{R\lambda/100}{1-\lambda}\right)}^{t}}+\frac{100}{{\left(1+\frac{r+(1-R/100)\lambda}{1-\lambda}+\frac{R\lambda/100}{1-\lambda}\right)}^{T}} \cr
 &= (C_{par}+\Delta)\sum_{t=1}^{T}\frac{{(1-\lambda)}^{t}}{{(1+r)}^{t}}+100\frac{{(1-\lambda)}^{T}}{{(1+r)}^{T}},
\end{split}
\end{equation}
which is smaller than the left-hand side of Eq.~\eqref{eq:ytm_mmI_proof_prop1_eq1}, unless $R=0$ or $\lambda=0$. It follows that it needs $\delta<\delta_{\sup}$ for Eq.~\eqref{eq:ytm_mmI_proof_prop1_eq1} to be satisfied (except when $R=0$ or $\lambda=0$, which both imply $\delta=\delta_{\sup}=0$). \newline
Given that the first part of the proof focusses on the case in which there is a change in the level of the coupon with respect to $C_{par}$, to complete the proof one needs to see what happens to the yield-to-maturity of a bond when to a first coupon increment $\Delta_{1}>0$ one adds a further increment $\Delta_{2}>0$.%
\footnote{The proof for negative variations is analogous.}
In this case one has to show that the yield-to-maturity $c_{par}+\delta_{1}$ (related to the bond with coupon $C_{par}+\Delta_{1}$) increases of a positive value $\delta_{2}$ implicitly defined by
\begin{equation}\label{eq:ytm_mmI_proof_prop1_eq6}
\begin{split}
100+(\Delta_{1}+\Delta_{2})\sum_{t=1}^{T} & \frac{{(1-\lambda)}^{t}}{{(1+r)}^{t}} = \cr
 &= \sum_{t=1}^{T}\frac{C_{par}+\Delta_{1}+\Delta_{2}}{{(1+c_{par}+\delta_{1}+\delta_{2})}^{t}}+\frac{100}{{(1+c_{par}+\delta_{1}+\delta_{2})}^{T}}.
\end{split}
\end{equation}
When $\delta_{2}=0$, the right-hand side of Eq.~\eqref{eq:ytm_mmI_proof_prop1_eq6} becomes, using Eqs.~\eqref{eq:ytm_mmI_proof_prop1_eq2} and~\eqref{eq:ytm_mmI_cpar},
\begin{equation}\label{eq:ytm_mmI_proof_prop1_eq7}
\begin{split}
\sum_{t=1}^{T} & \frac{C_{par}+\Delta_{1}+\Delta_{2}}{{(1+c_{par}+\delta_{1})}^{t}}+\frac{100}{{(1+c_{par}+\delta_{1})}^{T}}= \cr
 &= 100+\Delta_{1}\sum_{t=1}^{T}\frac{{(1-\lambda)}^{t}}{{(1+r)}^{t}}+\Delta_{2}\sum_{t=1}^{T}\frac{1}{{(1+c_{par}+\delta_{1})}^{t}} \cr
 &= 100+\Delta_{1}\sum_{t=1}^{T}\frac{{(1-\lambda)}^{t}}{{(1+r)}^{t}}+\Delta_{2}\sum_{t=1}^{T}{\left(\frac{1-\lambda}{1+r-R\lambda/100+\delta_{1}(1-\lambda)}\right)}^{t},
\end{split}
\end{equation}
which is greater than the left-hand side of Eq.~\eqref{eq:ytm_mmI_proof_prop1_eq6}, as $\delta_{1}<\delta_{\sup}=(R\lambda/100)/(1-\lambda)$. It follows that $\delta_{2}$ has to be positive for Eq.~\eqref{eq:ytm_mmI_proof_prop1_eq6} to be satisfied. \QED

\subsection{Proof of Proposition~\ref{th:ytm_mmI_prop2}}
\setcounter{equation}{0}

To prove that coupons greater than $C_{par}$ imply increasing yield-to-maturity term structures,%
\footnote{The proof for coupons smaller than $C_{par}$ and decreasing yield-to-maturity term structures is similar.}
one has to show that, for a bond with maturity $T+1$ and coupon $C_{par}+\Delta$, with $\Delta>0$, the parameter $\delta=\delta_{T+1}-\delta_{T}$ implicitly defined by the following equation, derived from Eqs.~\eqref{eq:ytm_mmI_ytm} and~\eqref{eq:ytm_mmI_paryield}, is positive:
\begin{equation}\label{eq:ytm_mmI_proof_prop2_eq1}
100+\Delta\sum_{t=1}^{T+1}\frac{{(1-\lambda)}^{t}}{{(1+r)}^{t}}=\sum_{t=1}^{T+1}\frac{C_{par}+\Delta}{{(1+c_{par}+\delta_{T}+\delta)}^{t}}+\frac{100}{{(1+c_{par}+\delta_{T}+\delta)}^{T+1}}.
\end{equation}
The parameters $\delta_{T+1}$ and $\delta_{T}$ are the differences with respect to $c_{par}$ of the yields-to-maturity of bonds with maturities $T+1$ and $T$, respectively, and coupons equal to $C_{par}+\Delta$. \newline
Assuming $\delta\leq 0\Leftrightarrow\delta_{T+1}\leq\delta_{T}$ and using Eq.~\eqref{eq:ytm_mmI_proof_prop1_eq2}, one has
\begin{equation}\label{eq:ytm_mmI_proof_prop2_eq2}
\begin{split}
100+\Delta\sum_{t=1}^{T+1}\frac{{(1-\lambda)}^{t}}{{(1+r)}^{t}} &= \sum_{t=1}^{T+1}\frac{C_{par}+\Delta}{{(1+c_{par}+\delta_{T+1})}^{t}}+\frac{100}{{(1+c_{par}+\delta_{T+1})}^{T+1}} \cr
 &\geq \sum_{t=1}^{T+1}\frac{C_{par}+\Delta}{{(1+c_{par}+\delta_{T})}^{t}}+\frac{100}{{(1+c_{par}+\delta_{T})}^{T+1}} \cr
 &\geq 100+\Delta\sum_{t=1}^{T}\frac{{(1-\lambda)}^{t}}{{(1+r)}^{t}}-\frac{100}{{(1+c_{par}+\delta_{T})}^{T}} \cr
 &\qquad +\frac{C_{par}+\Delta}{{(1+c_{par}+\delta_{T})}^{T+1}}+\frac{100}{{(1+c_{par}+\delta_{T})}^{T+1}},
\end{split}
\end{equation}
or
\begin{equation}\label{eq:ytm_mmI_proof_prop2_eq3}
\Delta\frac{{(1-\lambda)}^{T+1}}{{(1+r)}^{T+1}}\geq\frac{C_{par}+\Delta+100}{{(1+c_{par}+\delta_{T})}^{T+1}}-\frac{100}{{(1+c_{par}+\delta_{T})}^{T}}.
\end{equation}
Hence,
\begin{equation}\label{eq:ytm_mmI_proof_prop2_eq4}
\Delta\alpha_{T}+100\delta_{T}\geq\Delta,\quad\text{where}\quad\alpha_{T}:=\frac{{(1-\lambda)}^{T+1}}{{(1+r)}^{T+1}}{\big(1+c_{par}+\delta_{T}\big)}^{T+1}.
\end{equation}
It can be verified that $\alpha_{T}<1$, using the fact that $1+c_{par}+\delta_{\sup}=(1+r)/(1-\lambda)$ (see Eqs.~\ref{eq:ytm_mmI_cpar} and~\ref{eq:ytm_mmI_proof_prop1_eq4}), as
\begin{equation}\label{eq:ytm_mmI_proof_prop2_eq5}
\frac{{(1-\lambda)}^{T+1}}{{(1+r)}^{T+1}}{\big(1+c_{par}+\delta_{T}\big)}^{T+1}<\frac{{(1-\lambda)}^{T+1}}{{(1+r)}^{T+1}}{\big(1+c_{par}+\delta_{\sup}\big)}^{T+1}=1.
\end{equation}
Plugging the value of $\Delta$ given by Eq.~\eqref{eq:ytm_mmI_proof_prop2_eq4} in the right-hand side of Eq.~\eqref{eq:ytm_mmI_proof_prop1_eq2} one obtains
\begin{equation}\label{eq:ytm_mmI_proof_prop2_eq6}
100+\Delta\sum_{t=1}^{T}\frac{{(1-\lambda)}^{t}}{{(1+r)}^{t}}\leq\sum_{t=1}^{T}\frac{C_{par}+\Delta\alpha_{T}+100\delta_{T}}{{(1+c_{par}+\delta_{T})}^{t}}+\frac{100}{{(1+c_{par}+\delta_{T})}^{T}},
\end{equation}
which, using Eq.~\eqref{eq:ytm_mmI_generic} and dividing by $\Delta$, is equivalent to
\begin{equation}\label{eq:ytm_mmI_proof_prop2_eq7}
\sum_{t=1}^{T}\frac{{(1-\lambda)}^{t}}{{(1+r)}^{t}}\leq\alpha_{T}\!\sum_{t=1}^{T}\frac{1}{{\big(1+c_{par}+\delta_{T}\big)}^{t}}.
\end{equation}
It can be shown that the right-hand side of Eq.~\eqref{eq:ytm_mmI_proof_prop2_eq7} is increasing in $\delta_{T}$ by computing the corresponding derivative,
\begin{equation}\label{eq:ytm_mmI_proof_prop2_eq8}
\begin{split}
\frac{\partial}{\partial\delta_{T}} & {\left(\alpha_{T}\sum_{t=1}^{T}\frac{1}{{\big(1+c_{par}+\delta_{T}\big)}^{t}}\right)}= \cr
 &= \frac{{(1-\lambda)}^{T+1}}{{(1+r)}^{T+1}}{\left(\frac{\partial}{\partial\delta_{T}}\sum_{t=1}^{T}{\big(1+c_{par}+\delta_{T}\big)}^{T-t}\right)} \cr
 &= \frac{{(1-\lambda)}^{T+1}}{{(1+r)}^{T+1}}{\big(1+c_{par}+\delta_{T}\big)}^{T-1}\sum_{t=1}^{T}\frac{T-t}{{(1+c_{par}+\delta_{T})}^{t}},
\end{split}
\end{equation}
and observing that it is positive as long as $T\geq 2$ (which is the minimum number of maturities for which a term structure makes sense). \newline
It follows that, replacing $\alpha_{T}$ by $1$ and $\delta_{T}$ by $\delta_{\sup}$ in the right-hand side of Eq.~\eqref{eq:ytm_mmI_proof_prop2_eq7} and using the value of $c_{par}$ (see Eq.~\ref{eq:ytm_mmI_cpar}) and $\delta_{\sup}$ (see Eq.~\ref{eq:ytm_mmI_proof_prop1_eq4}), one obtains the following inequality,
\begin{equation}\label{eq:ytm_mmI_proof_prop2_eq9}
\alpha_{T}\sum_{t=1}^{T}\frac{1}{{\big(1+c_{par}+\delta_{T}\big)}^{t}}<\sum_{t=1}^{T}\frac{1}{{\big(1+c_{par}+\delta_{\sup}\big)}^{t}}=\sum_{t=1}^{T}\frac{{(1-\lambda)}^{t}}{{(1+r)}^{t}},
\end{equation}
which contradicts Eq.~\eqref{eq:ytm_mmI_proof_prop2_eq7}, and also the initial assumption that $\delta\leq 0$. \newline
To complete the proof, one can show that coupons higher (lower) than $C_{par}$ imply increasing (decreasing) bond price term structures, with the other factors being the same. From Eq.~\eqref{eq:ytm_mmI_bond}, one can write the price of a bond with maturity $T+1$ in terms of the price of a bond with maturity $T$ as
\begin{equation}\label{eq:ytm_mmI_proof_prop2_eq10}
\begin{split}
P_{T+1} &= P_{T}+\left(C\frac{{(1-\lambda)}^{T+1}}{{(1+r)}^{T+1}}+R\frac{\lambda{(1-\lambda)}^{T}}{{(1+r)}^{T+1}}+100\frac{{(1-\lambda)}^{T+1}}{{(1+r)}^{T+1}}-100\frac{{(1-\lambda)}^{T}}{{(1+r)}^{T}}\right) \cr
 &= P_{T}+\frac{{(1-\lambda)}^{T}}{{(1+r)}^{T}}\left(C\frac{1-\lambda}{1+r}+R\frac{\lambda}{1+r}+100\frac{1-\lambda}{1+r}-100\right).
\end{split}
\end{equation}
In can be noticed that the second term on the right-hand side of Eq.~\eqref{eq:ytm_mmI_proof_prop2_eq10} is equal to zero when $C=C_{par}$ and is greater (smaller) than zero when $C>C_{par}$ ($C<C_{par}$), so that
\begin{equation}\label{eq:ytm_mmI_proof_prop2_eq11}
\begin{array}{ll}
100<P_{1}<P_{2}<\cdots, & \text{if $C>C_{par}$} \cr
100>P_{1}>P_{2}<\cdots, & \text{if $C<C_{par}$}.
\end{array}
\end{equation}
One can thus conclude that the term structure of bond prices is increasing when the coupons are higher than $C_{par}$, and vice versa. \QED

\subsection{Proof of Remark~\ref{th:ytm_mmI_remark}}
\setcounter{equation}{0}

Using Eqs.~\eqref{eq:ytm_mmI_ytm} and Eq.~\eqref{eq:ytm_mmI_paryield}, it can be verified that for any bond with maturity $T$, coupons equal to $C_{par}+\Delta$, and related yield-to-maturity $c_{par}+\delta_{T}$, the following equation holds:
\begin{equation}\label{eq:ytm_mmI_proof_remark}
100+\Delta \sum_{t=1}^{T}\frac{{(1-\lambda)}^{t}}{{(1+r)}^{t}}=\sum_{t=1}^{T}\frac{C_{par}+\Delta}{{(1+c_{par}+\delta_{T})}^{t}}+\frac{100}{{(1+c_{par}+\delta_{T})}^{T}}.
\end{equation}
The right-hand side of Eq.~\eqref{eq:ytm_mmI_proof_remark} is equal to 100 when $\delta_{T}=\Delta/100$, because of Eq.~\eqref{eq:ytm_mmI_generic}, and is smaller (greater) than the left-hand side if $\Delta$ is positive (negative). Hence, Eq.~\eqref{eq:ytm_mmI_proof_remark} can be satisfied only if $|\delta_{T}|<|\Delta|/100$. \QED

\subsection{Proof of Proposition~\ref{th:ytm_mmII_prop1}}
\setcounter{equation}{0}

The proof follows the three steps described in the proof of Proposition~\ref{th:ytm_mmI_prop2}. \newline
\emph{Step 1}. It can be verified that, for any bond with coupon rate $c=C/100$, one has
\begin{equation}\label{eq:ytm_mmII_generic}
100=C\!\int_{0}^{T}e^{-ct}\dt+100\,e^{-cT}.
\end{equation}
In particular, the following equation holds when $c=c_{par}=C_{par}/100$: 
\begin{equation}\label{eq:ytm_mmII_generic_paryield}
100=C_{par}\!\int_{0}^{T}e^{-c_{par}t}\dt+100\,e^{-c_{par}T}.
\end{equation}
Lemma~\ref{th:ytm_mmII_lemma} and Eq.~\eqref{eq:ytm_mmII_generic_paryield} imply that the yield-to-maturity of any bond with maturity $T$ is equal to the par yield when the coupons are equal to $C_{par}$. Moreover, it follows from Eqs.~\eqref{eq:ytm_mmII_ytm} and~\eqref{eq:ytm_mmII_paryield} that when a bond has a coupon which differs from $C_{par}$ by $\Delta$, the variation $\delta$ of its yield-to-maturity with respect to $c_{par}$ is implicitly defined by
\begin{equation}\label{eq:ytm_mmII_proof_prop1_eq1}
\begin{split}
(C_{par}+\Delta)\!\int_{0}^{T}\,e^{-(r+\lambda) t}\dt &+ R\lambda\!\int_{0}^{T}e^{-(r+\lambda)t}\dt+100\,e^{-(r+\lambda)T}= \cr
 &= (C_{par}+\Delta)\!\int_{0}^{T}e^{-(c_{par}+\delta)t}\dt+100\,e^{-(c_{par}+\delta)T},
\end{split}
\end{equation}
or
\begin{equation}\label{eq:ytm_mmII_proof_prop1_eq2}
100+\Delta\!\int_{0}^{T}e^{-(r+\lambda)t}\dt=(C_{par}+\Delta)\!\int_{0}^{T}e^{-(c_{par}+\delta)t}\dt+100\,e^{-(c_{par}+\delta)T}.
\end{equation}
When $\delta=0$, the right-hand side of Eq.~\eqref{eq:ytm_mmII_proof_prop1_eq2} becomes, using Eqs.~\eqref{eq:ytm_mmII_generic_paryield} and~\eqref{eq:ytm_mmII_cpar},
\begin{equation}\label{eq:ytm_mmII_proof_prop1_eq3}
\begin{split}
(C_{par}+\Delta)\!\int_{0}^{T}e^{-(c_{par}+\delta)t}\dt+100\,e^{-(c_{par}+\delta)T} &= 100+\Delta\!\int_{0}^{T}e^{-c_{par}t}\dt \cr
 &= 100+\Delta\!\int_{0}^{T}e^{-\big(r+\lambda-\frac{R\lambda}{100}\big)t}\dt, 
\end{split}
\end{equation}
which is greater (lower) than the left-hand side of Eq.~\eqref{eq:ytm_mmII_proof_prop1_eq2} when $\Delta$ is positive (negative). It follows that $\delta$ has to be positive (negative) when $\Delta$ is positive (negative) for Eq.~\eqref{eq:ytm_mmII_proof_prop1_eq2} to be satisfied. \newline
\emph{Step 2}. The yield-to-maturity has an upper bound, which is independent of the coupon rate, given by $c_{par}+\delta_{\sup}$, where
\begin{equation}\label{eq:ytm_mmII_proof_prop1_eq4}
\delta_{\sup}=\frac{R\lambda}{100}.
\end{equation}
In fact, when $\delta=\delta_{\sup}$ the right-hand side of Eq.~\eqref{eq:ytm_mmII_proof_prop1_eq1} is equal to
\begin{equation}\label{eq:ytm_mmII_proof_prop1_eq5}
\begin{split}
(C_{par}+\Delta)\!\int_{0}^{T} & e^{-(c_{par}+\delta_{\sup})t}\dt+100\,e^{-(c_{par}+\delta_{\sup})T}= \cr
 &= (C_{par}+\Delta)\!\int_{0}^{T}e^{-\left(c_{par}+\frac{R\lambda}{100}\right)t}\dt+100\,e^{-\left(c_{par}+\frac{R\lambda}{100}\right)T} \cr
 &= (C_{par}+\Delta)\!\int_{0}^{T}e^{-(r+\lambda)t}\dt+100\,e^{-(r+\lambda)T},
\end{split}
\end{equation}
which is smaller than the left-hand side of Eq.~\eqref{eq:ytm_mmII_proof_prop1_eq1}, unless $R=0$ or $\lambda=0$. It follows that it needs $\delta<\delta_{\sup}$ for Eq.~\eqref{eq:ytm_mmII_proof_prop1_eq1} to be satisfied (except when $R=0$ or $\lambda=0$, which both imply $\delta=\delta_{\sup}=0$). \newline
\emph{Step 3}. To complete the proof, one has to see what happens to the yield-to-maturity of a bond when to a first coupon increment $\Delta_{1}>0$ one adds a further increment $\Delta_{2}>0$.%
\footnote{The proof for negative variations is analogous.}
In this case one has to show that the yield-to-maturity $c_{par}+\delta_{1}$ (related to the bond with coupon $C_{par}+\Delta_{1}$) increases of a positive value $\delta_{2}$ implicitly defined by
\begin{equation}\label{eq:ytm_mmII_proof_prop1_eq6}
\begin{split}
100+(\Delta_{1} &+ \Delta_{2})\!\int_{0}^{T}e^{-(r+\lambda)t}\dt \cr
&=(C_{par}+\Delta_{1}+\Delta_{2})\!\int_{0}^{T}e^{-(c_{par}+\delta_{1}+\delta_{2})t}\dt+100\,e^{-(c_{par}+\delta_{1}+\delta_{2})T}.
\end{split}
\end{equation}
When $\delta_{2}=0$, the right-hand side of Eq.~\eqref{eq:ytm_mmII_proof_prop1_eq6} becomes, using Eqs.~\eqref{eq:ytm_mmII_proof_prop1_eq2} and~\eqref{eq:ytm_mmII_cpar},
\begin{equation}\label{eq:ytm_mmII_proof_prop1_eq7}
\begin{split}
(C_{par}+\Delta_{1}+\Delta_{2})\!\int_{0}^{T} & e^{-(c_{par}+\delta_{1})t}\dt+100\,e^{-(c_{par}+\delta_{1})T}= \cr
 &= 100+\Delta_{1}\!\int_{0}^{T}e^{-(r+\lambda)t}\dt+\Delta_{2}\!\int_{0}^{T}e^{-(c_{par}+\delta_{1})t}\dt \cr
 &= 100+\Delta_{1}\!\int_{0}^{T}e^{-(r+\lambda)t}\dt+\Delta_{2}\!\int_{0}^{T}e^{-\left(r+\lambda-\frac{R\lambda}{100}+\delta_{1}\right)t}\dt,
\end{split}
\end{equation}
which is greater than the left-hand side of Eq.~\eqref{eq:ytm_mmII_proof_prop1_eq6}, as $\delta_{1}<\delta_{\sup}=R\lambda/100$. It follows that $\delta_{2}$ has to be positive for Eq.~\eqref{eq:ytm_mmII_proof_prop1_eq6} to be satisfied. \QED

\subsection{Proof of Proposition~\ref{th:ytm_mmII_prop2}}
\setcounter{equation}{0}

To prove that coupons greater than $C_{par}$ imply increasing yield-to-maturity term structures,%
\footnote{The proof for coupons smaller than $C_{par}$ and decreasing yield-to-maturity term structures is similar.}
one has to show that, for a bond with maturity $T+\epsilon$ and coupon $C_{par}+\Delta$, with $\Delta>0$, the parameter $\delta=\delta_{T+\epsilon}-\delta_{T}$ implicitly defined by the following equation, derived from Eqs.~\eqref{eq:ytm_mmII_ytm} and~\eqref{eq:ytm_mmII_paryield}, is positive:
\begin{equation}\label{eq:ytm_mmII_proof_prop2_eq1}
\begin{split}
100+\Delta & \int_{0}^{T+\epsilon}e^{-(r+\lambda)t}\dt \cr
&=(C_{par }+\Delta)\!\int_{0}^{T+\epsilon} e^{-(c_{par}+\delta_{T}+\delta)t}\dt+100\,e^{-(c_{par}+\delta_{T}+\delta)(T+\epsilon)}.
\end{split}
\end{equation}
The parameters $\delta_{T+\epsilon}$ and $\delta_{T}$ (for any $\epsilon>0$ arbitrary small) are the differences with respect to $c_{par}$ of the yields-to-maturity of bonds with maturities $T+\epsilon$ and $T$, respectively, and coupons equal to $C_{par}+\Delta$. \newline
Assuming $\delta\leq 0\Leftrightarrow\delta_{T+\epsilon}\leq\delta_{T}$ and using Eq.~\eqref{eq:ytm_mmII_proof_prop1_eq2} one has
\begin{equation}\label{eq:ytm_mmII_proof_prop2_eq2}
\begin{split}
100+\Delta\!\int_{0}^{T+\epsilon} & e^{-(r+\lambda)t}\dt= \cr
 &= (C_{par}+\Delta)\!\int_{0}^{T+\epsilon}e^{-(c_{par}+\delta_{T+\epsilon})t}\dt+100\,e^{-(c_{par}+\delta_{T+\epsilon})(T+\epsilon)} \cr
 &\geq (C_{par}+\Delta)\!\int_{0}^{T+\epsilon}e^{-(c_{par}+\delta_{T})t}\dt+100\,e^{-(c_{par}+\delta_{T})(T+\epsilon)} \cr
 &\geq 100+\Delta\!\int_{0}^{T}e^{-(r+\lambda)t}\dt-100\,e^{-(c_{par}+\delta_{T})T} \cr
 &\qquad+ (C_{par}+\Delta)\!\int_{T}^{T+\epsilon}e^{-(c_{par}+\delta_{T})t}\dt \cr
 &\qquad+ 100\,e^{-(c_{par}+\delta_{T})(T+\epsilon)}, \cr
\end{split}
\end{equation}
or
\begin{equation}\label{eq:ytm_mmII_proof_prop2_eq3}
\begin{split}
\Delta\!\int_{T}^{T+\epsilon}e^{-(r+\lambda)t}\dt &\geq (C_{par}+\Delta)\!\int_{T}^{T+\epsilon}e^{-(c_{par}+\delta_{T})t} \cr
 &\qquad- 100(c_{par}+\delta_{T})\int_{T}^{T+\epsilon}e^{-(c_{par}+\delta_{T})t} \cr
 &\geq (\Delta-100\delta_{T})\int_{T}^{T+\epsilon}e^{-(c_{par}+\delta_{T})t}.
\end{split}
\end{equation}
Hence,
\begin{equation}\label{eq:ytm_mmII_proof_prop2_eq4}
\Delta\alpha_{T}+100\delta_{T}\geq\Delta,\quad\text{where}\quad\alpha_{T}:=\frac{\int_{T}^{T+\epsilon}e^{-(r+\lambda)t}\dt}{\int_{T}^{T+\epsilon}e^{-(c_{par}+\delta_{T})t}\dt}.
\end{equation}
It can be verified that $\alpha_{T}<1$, using the fact that $c_{par}+\delta_{\sup}=r+\lambda$ (see Eqs.~\ref{eq:ytm_mmII_cpar} and~\ref{eq:ytm_mmII_proof_prop1_eq4}), as
\begin{equation}\label{eq:ytm_mmII_proof_prop2_eq5}
\frac{\int_{T}^{T+\epsilon}e^{-(r+\lambda)t}\dt}{\int_{T}^{T+\epsilon}e^{-(c_{par}+\delta_{T})t}\dt}<\frac{\int_{T}^{T+\epsilon}e^{-(r+\lambda)t}\dt}{\int_{T}^{T+\epsilon}e^{-(c_{par}+\delta_{\sup})t}\dt}=1.
\end{equation}
Plugging the value of $\Delta$ given by Eq.~\eqref{eq:ytm_mmII_proof_prop2_eq4} in the right-hand side of Eq.~\eqref{eq:ytm_mmII_proof_prop1_eq2} one obtains
\begin{equation}\label{eq:ytm_mmII_proof_prop2_eq6}
\begin{split}
100+\Delta\!\int_{0}^{T}e^{-(r+\lambda)t}\dt &\leq (C_{par}+\Delta\alpha_{T}+100\delta_{T})\!\int_{0}^{T}e^{-(c_{par}+\delta_{T})t}\dt \cr
 &\qquad+ 100\,e^{-(c_{par}+\delta_{T})T},
\end{split}
\end{equation}
which, using Eq.~\eqref{eq:ytm_mmII_generic} and dividing by $\Delta$, is equivalent to
\begin{equation}\label{eq:ytm_mmII_proof_prop2_eq7}
\int_{0}^{T}e^{-(r+\lambda)t}\dt\leq\alpha_{T}\!\int_{0}^{T}e^{-(c_{par}+\delta_{T})t}\dt.
\end{equation}
It can be shown that the right-hand side in Eq.~\eqref{eq:ytm_mmII_proof_prop2_eq7} is increasing in $\delta_{T}$ by computing the corresponding derivative,
\begin{equation}\label{eq:ytm_mmII_proof_prop2_eq8}
\begin{split}
\frac{\partial}{\partial\delta_{T}} & {\left(\alpha_{T}\!\int_{0}^{T}e^{-(c_{par}+\delta_{T})t}\dt\right)}= \cr
 &= {\left(\int_{T}^{T+\epsilon}e^{-(r+\lambda)t}\dt\right)}{\left(\frac{\partial}{\partial\delta_{T}}\frac{\int_{0}^{T}e^{-(c_{par}+\delta_{T})t}\dt}{\int_{T}^{T+\epsilon}e^{-(c_{par}+\delta_{T})t}\dt}\right)} \cr
 &= {\left(\int_{T}^{T+\epsilon}e^{-(r+\lambda)t}\dt\right)}\cdot \cr
 &\qquad\cdot {\left(\frac{e^{(c_{par}+\delta_{T})(T+\epsilon)}\big(\epsilon e^{-(c_{par}+\delta_{T})T}+Te^{(c_{par}+\delta_{T})\epsilon}-T-\epsilon\big)}{{\left(e^{(c_{par}+\delta_{T})\epsilon}-1\right)}^2}\right)}.
\end{split}
\end{equation}
Denoting $w:=c_{par}+\delta_{T}$, the sign of Eq.~\eqref{eq:ytm_mmII_proof_prop2_eq8} is equal to that of the following quantity, taken from the numerator of the second term,
\begin{equation}\label{eq:ytm_mmII_proof_prop2_eq8bis}
\epsilon e^{-wT}+Te^{w\epsilon}-T-\epsilon.
\end{equation}
It can be shown that Eq.~\eqref{eq:ytm_mmII_proof_prop2_eq8bis} represents a positive quantity. In fact, it is equal to 0 when $w=0$ and is increasing in $w$, as it can be shown by computing the corresponding derivative,
\begin{equation}\label{eq:ytm_mmII_proof_prop2_eq8ter}
\frac{\partial}{\partial w}\big(\epsilon e^{-wT}+Te^{w\epsilon}-T-\epsilon\big)=T\epsilon{\left(1-e^{-w(T-\epsilon)}\right)},
\end{equation}
and observing that it is positive for any $T>\epsilon$. \newline
Since the right-hand side in Eq.~\eqref{eq:ytm_mmII_proof_prop2_eq7} is increasing in $\delta_{T}$,one can use the fact that $\alpha_{T}<1$ and $c_{par}+\delta_{\sup}=r+\lambda$ (see Eqs.~\ref{eq:ytm_mmII_cpar} and~\ref{eq:ytm_mmII_proof_prop1_eq4}) to write the following inequality,
\begin{equation}\label{eq:ytm_mmII_proof_prop2_eq9}
\alpha_{T}\!\int_{0}^{T}e^{-(c_{par}+\delta_{T})t}\dt<\int_{0}^{T}e^{-(c_{par}+\delta_{\sup})t}\dt=\int_{0}^{T}e^{-(r+\lambda)t}\dt,
\end{equation}
which contradicts Eq.~\eqref{eq:ytm_mmII_proof_prop2_eq7}, and also the initial assumption that $\delta\leq 0$. \newline
To complete the proof, one can show that coupons higher (lower) than $C_{par}$ imply increasing (decreasing) bond price term structures, with the other factors being the same. From Eq.~\eqref{eq:ytm_mmII_bond} one can write the price of a bond with maturity $T+\epsilon$ in terms of the price of a bond with maturity $T$, as
\begin{equation}\label{eq:ytm_mmII_proof_prop2_eq10}
\begin{split}
P_{T+\epsilon} &= P_{T}+(C+R\lambda)\!\int_{T}^{T+\epsilon}e^{-(r+\lambda)t}\dt +100\,e^{-(r+\lambda)(T+\epsilon)}-100\,e^{-(r+\lambda)T} \cr
 &= P_{T}+\big(C+R\lambda-100(r+\lambda)\big).
\end{split}
\end{equation}
It can be noticed that the second term in the right-hand side of Eq.~\eqref{eq:ytm_mmII_proof_prop2_eq10} is equal to zero when $C=C_{par}$ and is greater (smaller) than zero when $C>C_{par}$ ($C<C_{par}$), so that
\begin{equation}\label{eq:ytm_mmII_proof_prop2_eq11}
\begin{array}{ll}
100<P_{t_{1}}<P_{t_{2}}<\cdots, & \text{if $C>C_{par}$} \cr
100>P_{t_{1}}>P_{t_{2}}<\cdots, & \text{if $C<C_{par}$},
\end{array}
\end{equation}
for $0<t_{1}<t_{2}<\cdots$. One can thus conclude that the term structure of bond prices is increasing when the coupons are higher than $C_{par}$, and vice versa. \QED

\subsection{Proof of Remark~\ref{th:ytm_mmII_remark}}
\setcounter{equation}{0}

Using Eqs.~\eqref{eq:ytm_mmII_ytm} and~\eqref{eq:ytm_mmII_paryield}, it can be verified that for any bond with maturity $T$, coupons equal to $C_{par}+\Delta$, and related yield-to-maturity $c_{par}+\delta_{T}$, the following equation holds:
\begin{equation}\label{eq:ytm_mmII_proof_remark}
100+\Delta\!\int_{0}^{T}e^{-(r+\lambda)t}\dt=(C_{par }+\Delta)\!\int_{0}^{T}e^{-(c_{par}+\delta_{T})t}\dt+100\,e^{-(c_{par}+\delta_{T})T}.
\end{equation}
The right-hand side of Eq.~\eqref{eq:ytm_mmII_proof_remark} is equal to 100 when $\delta_{T}=\Delta/100$, because of Eq.~\eqref{eq:ytm_mmII_generic}, and is smaller (greater) than the left-hand side if $\Delta$ is positive (negative). Hence, Eq.~\eqref{eq:ytm_mmII_proof_remark} can be satisfied only if $|\delta_{T}|<|\Delta|/100$. \QED

\subsection{The mathematics of CDS spreads}\label{sec:CDS}
\setcounter{equation}{0}

A credit default swap (CDS) is a derivative security that provides insurance against the risk of default of a particular company, known as reference entity. When a credit event occurs, the insurance buyer has the right to sell to the insurance seller bonds issued by the reference entity for their face value. In return, the protection buyer makes to the protection seller a periodic payment, known as CDS premium, until the maturity of the CDS or the credit event occurs, whichever comes first. The fair CDS premium is computed so that the expected discounted premium leg (i.e., the payments stream that the protection buyer expects to pay) and the expected discounted default leg (i.e., the payment that the protection seller expects to make in case of default) agree. Of course, both legs depend on the default probabilities, the recovery rate, and the discount rates.

\subsection*{Discrete time model}

In the discrete time model, the premium is assumed to be paid at times $t\in\{1,\dots\,T\}$, where $T$ is the CDS maturity. For the default time $\tau$ a geometric distribution is assumed and $\lambda(t)$ denotes the conditional probability of being in default at time $t$, given the survival up to time $t-1$. The probability of survival for $t$ years is thus given by
\begin{equation}
\bP(\tau>t)=\prod_{s=1}^{t}\big(1-\lambda(s)\big).
\end{equation}
As in Section~\ref{sec:ytm_mmI}, a constant risk-free rate is assumed.

If one also assumes that $\lambda(t)=\lambda$ is constant, the fair CDS spread can be written as
\begin{equation}
S_{T}=\frac{(1-R)\sum_{t=1}^T\frac{\lambda (1-\lambda)^{t-1}}{(1+r)^t}}{\sum_{t=1}^T\frac{(1-\lambda)^t}{(1+r)^t}}=(1-R)\frac{\lambda}{1-\lambda}
\end{equation}
that shows that the spread is constant in T.

On the other hand, if one assumes that $\lambda(t)=\lambda_{t}$ is piecewise constant
\begin{equation}
\lambda(t) =
\begin{cases}
\lambda_{1} & \textrm{for $0 < t \leq 1$} \cr
\lambda_{2} & \textrm{for $1 < t \leq 2$} \cr
\ldots \cr
\lambda_T & \textrm{for $T-1 < t \leq T$}
\end{cases}
\end{equation}
one can write the fair CDS spread as
\begin{equation}
S_{T}=\frac{(1-R)\sum_{t=1}^T{\left(\frac{\lambda_t}{(1+r)^t}\prod_{j=1}^{t-1}(1-\lambda_j)\right)}}{\sum_{t=1}^T\frac{1}{(1+r)^t}\prod_{j=1}^t(1-\lambda_j)}.
\end{equation}
One can easily see that $S_{T}$ is not constant in $T$ in this case. In fact, one has that $S_{2}>S_{1}$ if $\lambda_{2}>\lambda_{1}$, and $S_{2}<S_{1}$ if $\lambda_{2}<\lambda_{1}$, as can be seen from the following equation:
\begin{equation}
\begin{split}
S_{2}-S_{1} &= \frac{(1-R){\left(\frac{\lambda_{1}}{1+r}+\frac{\lambda_{2}(1-\lambda_{1})}{(1+r)^2}\right)}}{\frac{1-\lambda_{1}}{1+r}+\frac{(1-\lambda_{1})(1-\lambda_{2})}{(1+r)^2}}-\frac{(1-R)\frac{\lambda_{1}}{1+r}}{\frac{1-\lambda_{1}}{1+r}} \cr
 &= \frac{1-R}{1-\lambda_{1}}\frac{\lambda_{2}-\lambda_{1}}{2+r-\lambda_{2}}.
\end{split}
\end{equation}
Moreover, one has that $S_1<S_2<\ldots<S_{T}$ if $\lambda_{1}<\lambda_{2}<\ldots\lambda_T$, for any $T>2$, and vice versa.

However, if $\lambda_t$ is not monotonic, nothing can be said on the shape of the CDS curve in general. This result can already be seen if one considers the difference between the 3-year spread and the 2-year spread,
\begin{equation}
\begin{split}
S_{3}-S_{2}= & \frac{(1-R){\left(\frac{\lambda_{1}}{1+r}+\frac{\lambda_{2}(1-\lambda_{1})}{(1+r)^2}+\frac{\lambda_3(1-\lambda_{1})(1-\lambda_{2})}{(1+r)^3}\right)}}{\frac{1-\lambda_{1}}{1+r}+\frac{(1-\lambda_{1})(1-\lambda_{2})}{(1+r)^2}+\frac{(1-\lambda_{1})(1-\lambda_{2})(1-\lambda_3)}{(1+r)^2}} \cr
 &\qquad -\frac{(1-R){\left(\frac{\lambda_{1}}{1+r}+\frac{\lambda_{2}(1-\lambda_{1})}{(1+r)^2}\right)}}{\frac{1-\lambda_{1}}{1+r}+\frac{(1-\lambda_{1})(1-\lambda_{2})}{(1+r)^2}},
\end{split}
\end{equation}
that can be positive or negative depending on the values of $\lambda_{1}$, $\lambda_{2}$, and $\lambda_3$.

\subsection*{Continuous time model}

In the continuous time model, a deterministic default intensity $\lambda(t)$ is assumed and the probability of survival for $t$ years is defined as
\begin{equation}
\bP(\tau>t)=e^{-\int_{0}^{t}\lambda(s)ds}.
\end{equation}
If one assumes that $\lambda(t)=\lambda$ is constant, the fair CDS spread can be written as
\begin{equation}
S_{T}=\frac{(1-R)\int_0^T \lambda e^{-(r+\lambda)s}ds}{\int_0^Te^{-(r+\lambda)s}ds}=(1-R)\lambda
\end{equation}
that shows that the spread is constant in $T$.

On the other hand, if one assumes that $\lambda(t)$ is continuous piecewise linear, then conclusions similar to the discrete time case can be inferred. In general one has
\begin{equation}
S_{T}=\frac{(1-R)\int_0^T \lambda(s) e^{-(r+\lambda(s))s}ds}{\int_{0}^{T}e^{-(r+\lambda(s))s}ds}
\end{equation}
If we assume that $\lambda(t)$ is continuous, we can use the fundamental theorem of the integral calculus to compute the derivative with respect to $T$:
\begin{equation}
\frac{\partial S_{T}}{\partial T}=(1-R)\frac{e^{-(r+\lambda(T))T}{\left(\lambda(T)\int_0^Te^{-(r+\lambda(s))s}ds-\int_0^T\lambda(s)e^{-(r+\lambda(s))s}ds\right)}}{\bigg(\int_0^Te^{-(r+\lambda(s))s}ds\bigg)^2}.
\end{equation}
To see if the spread is a monotonic function in $T$ it is sufficient to analyze the sign of  
\begin{equation}\label{eq:sign}
\lambda(T) \int_0^Te^{-(r+\lambda(s))s}ds-\int_0^T\lambda(s)e^{-(r+\lambda(s))s}ds,
\end{equation}
being all the other quantities positive.

Under the assumption of a piecewise linear function, we can for example write an increasing $\lambda(t)$ as
\begin{equation}
\lambda(t) =
\begin{cases}
\alpha + \beta_1 t & \textrm{for $0 \leq t \leq T_{1}$} \cr
\alpha + \beta_1 t + \beta_2 (t-T_{1}) & \textrm{for $T_{1} < t \leq T_{2}$}
\end{cases}
\end{equation}
where $T_{1}<T_{2}$, and $\alpha$, $\beta_1$ and $\beta_2$ are positive constant, or a function that is increasing up to the time $T_{1}$ and then decreasing as
\begin{equation}
\lambda(t) =
\begin{cases}
\alpha + \beta_1 t & \text{for $0 \leq t \leq T_{1}$} \cr
\alpha + \beta_1 t - \beta_2 (t-T_{1}) & \text{for $T_{1} < t \leq T_{2}$}
\end{cases}
\end{equation}

In the two different cases we have respectively that the quantity defined in \eqref{eq:sign} is given by
\begin{equation}
\begin{cases}
(\alpha+\beta_1 T){\left(\int_0^T e^{-(r+\alpha +\beta_1 s)s}ds\right)}-\int_0^T(\alpha+\beta_1 s)e^{-(r+\alpha+\beta_1 s)s}ds & \text{for $T \leq T_{1}$} \cr
(\alpha+\beta_1 T_{1}+\beta_2 (T-T_{1}))\cdot \cr
 \qquad \cdot\left(\int_0^{T_{1}}e^{-(r+\alpha +\beta_1 s)s}ds+\int_{T_{1}}^Te^{-(r+\alpha+\beta_1 T_{1}+\beta_2 (s-T_{1}))s}ds\right) \cr
 \qquad -\int_0^{T_{1}}(\alpha+\beta_1 s)e^{-(r+\alpha+\beta_1 s)s}ds & \text{for $T_{1}<T\leq T_{2}$} \cr
 \qquad -\int_{T_{1}}^T(\alpha+\beta_1 T_{1}+\beta_2 (s-T_{1}))e^{-(r+\alpha+\beta_1 T_{1}+\beta_2 (s-T_{1}))s}ds
\end{cases}
\end{equation}

and

\begin{equation}
\begin{cases}
(\alpha+\beta_1 T){\left(\int_0^T e^{-(r+\alpha +\beta_1 s)s}ds\right)}-\int_0^T(\alpha+\beta_1 s)e^{-(r+\alpha+\beta_1 s)s}ds & \text{for $T \leq T_{1}$} \cr
(\alpha+\beta_1 T_{1}-\beta_2 (T-T_{1}))\cdot \cr
 \qquad \cdot{\left(\int_0^{T_{1}}e^{-(r+\alpha +\beta_1 s)s}ds+\int_{T_{1}}^Te^{-(r+\alpha+\beta_1 T_{1}-\beta_2 (s-T_{1}))s}ds\right)} \cr
 \qquad -\int_0^{T_{1}}(\alpha+\beta_1 s)e^{-(r+\alpha+\beta_1 s)s}ds & \text{for $T_{1}<T\leq T_{2}$} \cr
 \qquad -\int_{T_{1}}^T(\alpha+\beta_1 T_{1}-\beta_2 (s-T_{1}))e^{-(r+\alpha+\beta_1 T_{1}-\beta_2 (s-T_{1}))s}ds
\end{cases}
\end{equation}

We can easily see that the term for $T\leq T_{1}$ is equal in both cases and positive, while it would be negative if $\beta_1$ was negative (that is $\lambda(t)$ was decreasing for $t<T_{1}$). The term for $T>T_{1}$ is instead positive in the first case (would be negative if $\beta_1<0$ and $\beta_2<0$), while in the second case its sign depends on the value of the coefficients $\alpha$, $\beta_1$ and $\beta_2$, as well as on the value of $T$.
We can thus conclude that, while if $\lambda(t)$ is monotonic the spread is also monotonic, if $\lambda(t)$ is not monotonic we can't a priori say that the spread follows the same trend.

Finally if we assume that $\lambda(t)$ is piecewise constant, we have again results analogous to the discrete time case. 
Let us define
\begin{equation}
\lambda(t) =
\begin{cases}
\lambda_{1} & \textrm{for $t \leq 1$} \cr
\lambda_{2} & \textrm{for $1< t \leq 2$} \cr
\ldots \cr
\lambda_{T} & \textrm{for $T-1 < t \leq T$} \cr
\end{cases}
\end{equation}
we can write the CDS spread as
\begin{equation}
S_{T}=\frac{(1-R){\left(\int_0^1\lambda_{1} e^{-(r+\lambda_{1})s}ds+\int_{1}^2 \lambda_{2} e^{-(r+\lambda_{2})s}ds+ \ldots + \int_{T-1}^T \lambda_T e^{-(r+\lambda_T)s}ds\right)}}{\int_0^1 e^{-(r+\lambda_{1})s}ds+\int_{1}^2  e^{-(r+\lambda_{2})s}ds+ \ldots + \int_{T-1}^T  e^{-(r+\lambda_T)s}ds}.
\end{equation}
Clearly the spread depends on the value of $\lambda_i, i=1, \ldots, T$ and on the maturity $T$.
Let us consider for simplicity
\begin{equation}
\lambda(t) =
\begin{cases}
\lambda_{T_{1}} & \textrm{for $t \leq T_{1}$} \cr
\lambda_{T_{2}} & \textrm{for $T_{1}< t \leq T_{2}$} \cr
\lambda_{T_{3}} & \textrm{for $T_{2} < t \leq T_{3}$} \cr
\end{cases}
\end{equation}
with $T\leq T_{3}$,
we are interested in studying the sign of $\frac{\partial S_{T}}{\partial T}$ in the intervals. According to the maturity $T$ we have
\begin{equation}
S_{T} =
\begin{cases}
(1-R)\frac{\int_0^T\lambda_{T_{1}} e^{-(r+\lambda_{T_{1}})s}ds}{\int_0^T e^{-(r+\lambda_{T_{1}})s}ds} & \textrm{for $T \leq T_{1}$} \cr
(1-R)\frac{\int_0^{T_{1}}\lambda_{T_{1}} e^{-(r+\lambda_{T_{1}})s}ds+\int_{T_{1}}^T \lambda_{T_{2}} e^{-(r+\lambda_{T_{2}})s}ds}{\int_0^{T_{1}} e^{-(r+\lambda_{T_{1}})s}ds+\int_{T_{1}}^T  e^{-(r+\lambda_{T_{2}})s}ds} & \textrm{for $T_{1}< T \leq T_{2}$} \cr
(1-R)\frac{\int_0^{T_{1}}\lambda_{T_{1}} e^{-(r+\lambda_{T_{1}})s}ds+\int_{T_{1}}^{T_{2}} \lambda_{T_{2}} e^{-(r+\lambda_{T_{2}})s}ds+\int_{T_{2}}^T \lambda_{T_{3}} e^{-(r+\lambda_{T_{3}})s}ds}{\int_0^{T_{1}} e^{-(r+\lambda_{T_{1}})s}ds+\int_{T_{1}}^{T_{2}}  e^{-(r+\lambda_{T_{2}})s}ds+\int_{T_{2}}^T  e^{-(r+\lambda_{T_{3}})s}ds} & \textrm{for $T_{2} < T \leq T_{3}$}.\cr
\end{cases}
\end{equation}

Consequently we have that 
\begin{equation}
\frac{\partial S_{T}}{\partial T} =
\begin{cases}
0 & \textrm{for $T \leq T_{1}$} \cr
\frac{(1-R)e^{-(r+\lambda_{T_{2}})T}}{(\int_0^{T_{1}} e^{-(r+\lambda_{T_{1}})s}ds+\int_{T_{1}}^T  e^{-(r+\lambda_{T_{2}})s}ds)^2}\big(\int_0^{T_{1}} e^{-(r+\lambda_{T_{1}})s}ds (\lambda_{T_{2}}-\lambda_{T_{1}})\big) & \textrm{for $T_{1}< T \leq T_{2}$} \cr
\frac{(1-R)e^{-(r+\lambda_{T_{3}})T}}{(\int_0^{T_{1}} e^{-(r+\lambda_{T_{1}})s}ds+\int_{T_{1}}^{T_{2}}  e^{-(r+\lambda_{T_{2}})s}ds+\int_{T_{2}}^T  e^{-(r+\lambda_{T_{3}})s}ds)^2}\cdot \cr
\quad \cdot \big(\int_0^{T_{1}} e^{-(r+\lambda_{T_{1}})s}ds (\lambda_{T_{3}}-\lambda_{T_{1}})+\int_{T_{1}}^{T_{2}} e^{-(r+\lambda_{T_{2}})s}ds (\lambda_{T_{3}}-\lambda_{T_{2}})\big) & \textrm{for $T_{2} < T \leq T_{3}$}.\cr
\end{cases}
\end{equation}
Thus, it can be seen that
\begin{enumerate}
\item for $T\leq T_{1}$ the spread is constant in $T$, being $\lambda(t)=\lambda_{T_{1}}$ constant in the time interval;
\item for $T_{1}<T\leq T_{2}$, the spread is monotonic and has the same shape as $\lambda(t)$ (meaning that is increasing in $T$ if $\lambda_{T_{1}}<\lambda_{T_{2}}$ and vice versa);
\item for $T_{2}<T\leq T_{3}$, it depends on the specific values of $\lambda_{T_{1}},\lambda_{T_{2}}$, and $\lambda_{T_{3}}$. 
\end{enumerate}

\newpage

\section*{Tables and figures}
\addcontentsline{toc}{section}{Tables and figures}

\begin{table}[H]
\centering
\mycaption{Symbols used in the paper and their definitions}{}\label{tab:ytm_def}
\begin{tabularx}{\textwidth}{l >{\rule{0pt}{1em}}X<{\rule[-0.5em]{0pt}{0pt}}}
\toprule
Symbol & \multicolumn{1}{l}{Definition} \cr
\midrule
$r$ & Risk-free interest rate. The interest rate paid by a default-free bond ($r>0$). \cr
$T$ & Maturity. The date in which the nominal value of the bond is paid to the bondholder ($T\in\{1,2,\dots\}$ in the discrete-time model and $T>0$ in the continuous-time model). \cr
$C$ & Coupon value. The value of the coupons that are periodically paid by the bond ($C\geq 0$). \cr
$c$ & Coupon rate. The coupon value divided by the nominal value of the bond ($c=C/100$). \cr
$\lambda$ & Unconditional default probability (discrete-time model, $\lambda\in[0,1]$) or default arrival rate (continuous-time model, $\lambda\geq 0$). See Assumption~\ref{ass:ytm_mmI_defpr} on page~\pageref{ass:ytm_mmI_defpr} and Assumption~\ref{ass:ytm_mmII_defpr} on page~\pageref{ass:ytm_mmII_defpr}. \cr
$R$ & Recovery amount. The sum that the holder of a bond receives in case of default ($R\in[0,100]$). \cr
$R/100$ & Recovery rate. The recovery amount divided by the nominal value of the bond. \cr
$P_{T}$ & Bond price. Price at time 0 of a bond with maturity $T$. \cr
$\tau$ & Default time. Time at which the default of a bond happens ($\tau\in\{1,2,\dots\}$ in the discrete-time model and $\tau>0$ in the continuous-time model). \cr
$y_{T}$ & Yield-to-maturity of a bond with maturity $T$. See Definition~\ref{def:ytm_mmI_ytm} on page~\pageref{def:ytm_mmI_ytm}. \cr
$C_{par}$ & Par value of the coupon. See Definition~\ref{def:ytm_mmI_paryield} on page~\pageref{def:ytm_mmI_paryield}. \cr
$c_{par}$ & Par yield. The par value of the coupon divided by the nominal value of the bond ($c_{par}=C_{par}/100$). \cr
$\Delta$ & Difference between the coupon value $C$ and par value of the coupon $C_{par}$ of a bond. \cr
$\delta$ & Difference between the yield-to-maturity $y_{T}$ and par yield $c_{par}$ of a bond. \cr
$S_{T}$ & Fair CDS spread for a contract with maturity $T$. \cr
\bottomrule
\end{tabularx}
\end{table}

\newpage

\newpage


\addcontentsline{toc}{section}{References}
\bibliographystyle{}
\bibliography{}

\end{document}